\documentclass[useAMS,usenatbib]{mnras}
\usepackage{amssymb}
\usepackage{amsmath}
\usepackage{lipsum}
\usepackage{natbib}
\usepackage{multirow}% http://ctan.org/pkg/multirow
\usepackage{hhline}% http://ctan.org/pkg/hhline
\usepackage[pdftex]{graphicx}
\graphicspath{./figs/}

\newcommand{\be}{\begin{equation}}
\newcommand{\ee}{\end{equation}}

\RequirePackage{color}\definecolor{orange}{rgb}{1,0.5,0}
\newcommand\rev[1]{#1} %\textbf{#1}}

\newcommand{\xco}{$X_{\rm CO}$ }
\title[Kennicutt-Schmidt on FIRE]{What FIREs Up Star Formation: the Emergence of the Kennicutt-Schmidt Law from Feedback}
\author[M. E. Orr et al.]{Matthew E. Orr$^{1}$\thanks{E-mail:
meorr@caltech.edu}, Christopher C. Hayward$^{2,3,1}$, Philip F. Hopkins$^{1}$, \newauthor T. K. Chan$^4$, Claude-Andr\'e Faucher-Gigu\`{e}re$^{5}$, Robert Feldmann$^{6,7}$, \newauthor Du\v{s}an Kere\v{s}$^{4}$, Norman Murray$^8$, and Eliot Quataert$^{7}$\\\
$^{1}$TAPIR, Mailcode 350-17, California Institute of Technology, Pasadena, CA 91125, USA\\
$^{2}$Center for Computational Astrophysics, Flatiron Institute, 162 Fifth Avenue, New York, NY 10010, USA\\
$^{3}$Harvard-Smithsonian Center for Astrophysics, 60 Garden Street, Cambridge, MA 02138, USA\\
$^{4}$Department of Physics, Center for Astrophysics and Space Science, University of California at San Diego, 9500 Gilman Drive, \\ La Jolla, CA 92093, USA\\
$^{5}$Department of Physics and Astronomy and CIERA, Northwestern University, 2145 Sheridan Road, Evanston, IL 60208, USA\\
$^{6}$Institute for Computational Science, University of Zurich, Zurich CH-8057, Switzerland\\
$^{7}$Department of Astronomy and Theoretical Astrophysics Center, University of California, Berkeley, CA 94720-3411, USA\\
$^{8}$Canadian Institute for Theoretical Astrophysics, 60 St George Street, University of Toronto, ON M5S 3H8, Canada\\
}
\RequirePackage[normalem]{ulem} %DIF PREAMBLE
\RequirePackage{color}\definecolor{RED}{rgb}{1,0,0}\definecolor{BLUE}{rgb}{0,0,1} %DIF PREAMBLE
\begin{document}

\date{Draft date: \today}

\pagerange{\pageref{firstpage}--\pageref{lastpage}} \pubyear{2016}

\maketitle

\label{firstpage}

\begin{abstract}
We present an analysis of the global and spatially-resolved Kennicutt-Schmidt (KS) star formation relation in the FIRE (Feedback In Realistic Environments) suite of cosmological simulations, including halos with $z = 0$ masses ranging from $10^{10}$ -- $10^{13}$ M$_{\odot}$.  We show that the KS relation emerges and is robustly maintained due to the effects of feedback on local scales regulating star-forming gas, independent of the particular small-scale star formation prescriptions employed.  We demonstrate that the time-averaged KS relation is relatively independent of redshift and spatial averaging scale, and that the star formation rate surface density is weakly dependent on metallicity and inversely dependent on orbital dynamical time. At constant star formation rate surface density, the `Cold \& Dense' gas surface density (gas with $T < 300$~K and $n > 10$~cm$^{-3}$, used as a proxy for the molecular gas surface density) of the simulated galaxies is $\sim$0.5~dex less than observed at $\sim$kpc scales.  This discrepancy may arise from underestimates of the local column density at the particle-scale for the purposes of shielding in the simulations.  Finally, we show that on scales larger than individual giant molecular clouds, the primary condition that determines whether star formation occurs is whether a patch of the galactic disk is thermally Toomre-unstable (not whether it is self-shielding): once a patch can no longer be thermally stabilized against fragmentation, it collapses, becomes self-shielding, cools, and forms stars, regardless of epoch or environment.
\end{abstract}

\begin{keywords}
galaxies: star formation -- instabilities -- opacity -- galaxies: evolution, formation -- methods: numerical.
\end{keywords}

%------------------------------------------------------------------------------------
\section{Introduction}

Understanding star formation and its effects on galactic scales has been integral to assembling the story of the growth and subsequent evolution of the baryonic components of galaxies.  Observationally, the rate at which gas is converted into stars is characterized by the Kennicutt-Schmidt (KS) relation, which is a power law correlation between the star formation and gas surface densities in galaxies that holds over several orders of magnitude (\citealt{Schmidt1959, Kennicutt1998}; see \citealt{Kennicutt2012} for a recent review).

Numerous studies of the KS relation have shown that star formation is inefficient on galactic scales, with only a few per cent of a galaxy's gas mass being converted to stars per galactic free-fall time \citep{Kennicutt1998, Kennicutt2007, Daddi2010, Genzel2010}.  Understanding what regulates the efficiency of star formation and results in the observed KS relation is therefore key to understanding the formation and dynamics of galaxies.  Some authors \citep[e.g.][]{Thompson2005, Murray2010, Murray2011, Ostriker2011, Faucher-Giguere2013, Hayward2015, Semenov2016, Grudic2016} argue that star formation is \emph{locally} efficient, in the sense that tens of per cent of the mass of a gravitationally bound gas clump within a giant molecular cloud (GMC) can be converted into stars on the local free-fall time, and that local stellar feedback processes -- including supernovae (SNe), radiation pressure, photoheating and stellar winds -- must stabilize gas discs against catastrophic gravitational collapse, thereby resulting in the low global star formation efficiencies that are observed.  However, others claim on both theoretical and observational grounds that star formation is locally inefficient, with only of order a few per cent of the mass of clumps being converted into stars on a free-fall time independent of their density \citep{Padoan1995, Krumholz2007a, Lee2016}.

In either scenario, the KS law is considered to be an emergent relation that holds on galactic scales and results from a complex interplay of the physical processes that trigger star formation and those that regulate it.  It has also been argued and observed that the KS relation breaks down below some length- and time-scales \citep{Onodera2010, Schruba2010, Feldmann2011a, calzetti2012, Kruijssen2014}.  \citet{calzetti2012} found that the KS relation breaks down due to incomplete sampling of star-forming molecular clouds' mass function on length scales of less than $\sim 1$ kpc. \citet{Feldmann2012} claim that this breakdown on sub-kpc scales occurs due to the stochastic nature of star formation itself.  Furthermore, \citet{Kruijssen2014} argue that the various tracers of gas column density and star formation rate surface density require averaging over some spatial and temporal scales; consequently, when sufficiently small length scales are probed, a tight correlation between the star formation rate surface density and the gas column density should not be observed.  Understanding the scales where the KS law holds therefore informs our theories of star formation as well. 

On the length scales where the KS relation does hold, the canonical power law of the total gas relation is $\Sigma_{\rm SFR} \propto \Sigma_{\rm gas}^{1.4}$ with $\Sigma_{\rm SFR}$ being the star formation rate surface density and $\Sigma_{\rm gas}$ being the total gas surface density \citep{Kennicutt1998}.  However, there has been much debate regarding the power-law index of the relation and its physical origin; the previous literature has found KS relations ranging from highly sublinear to quadratic \citep{Bigiel2008, Daddi2010, Genzel2010, Feldmann2011b, Feldmann2012, Narayanan2012, Shetty2013, Shetty2014a, Shetty2014b, Becerra2014}.  Some of the disagreement owes to the particular formulation of the KS relation considered, such as whether $\Sigma_{\rm HI + H_2}$ (total atomic + molecular hydrogen column) or $\Sigma_{\rm H_2}$ (molecular column alone) is employed \citep[e.g. ][]{Rownd1999, Wong2002, Krumholz2007b}, with the $\Sigma_{\rm H_2}$ relation typically having slope $\sim 1$. The relation may in principle also depend on the star formation tracer used (e.g. H$\alpha$, far-infrared, or ultraviolet).  Furthermore, there are questions as to whether the index depends on spatial resolution -- even on scales larger than the length scale below which the relation fails altogether -- or if there are multiple tracks to the KS relation, each with different slopes across several decades in gas surface density \citep{Ostriker2010, Liu2011, Ostriker2011, Feldmann2011b, Feldmann2012, Faucher-Giguere2013}.
%%%%%%%%%%% FIGURE 1 - EXAMPLE MAP
\begin{figure}
	\centering
	\includegraphics[width=0.46\textwidth]{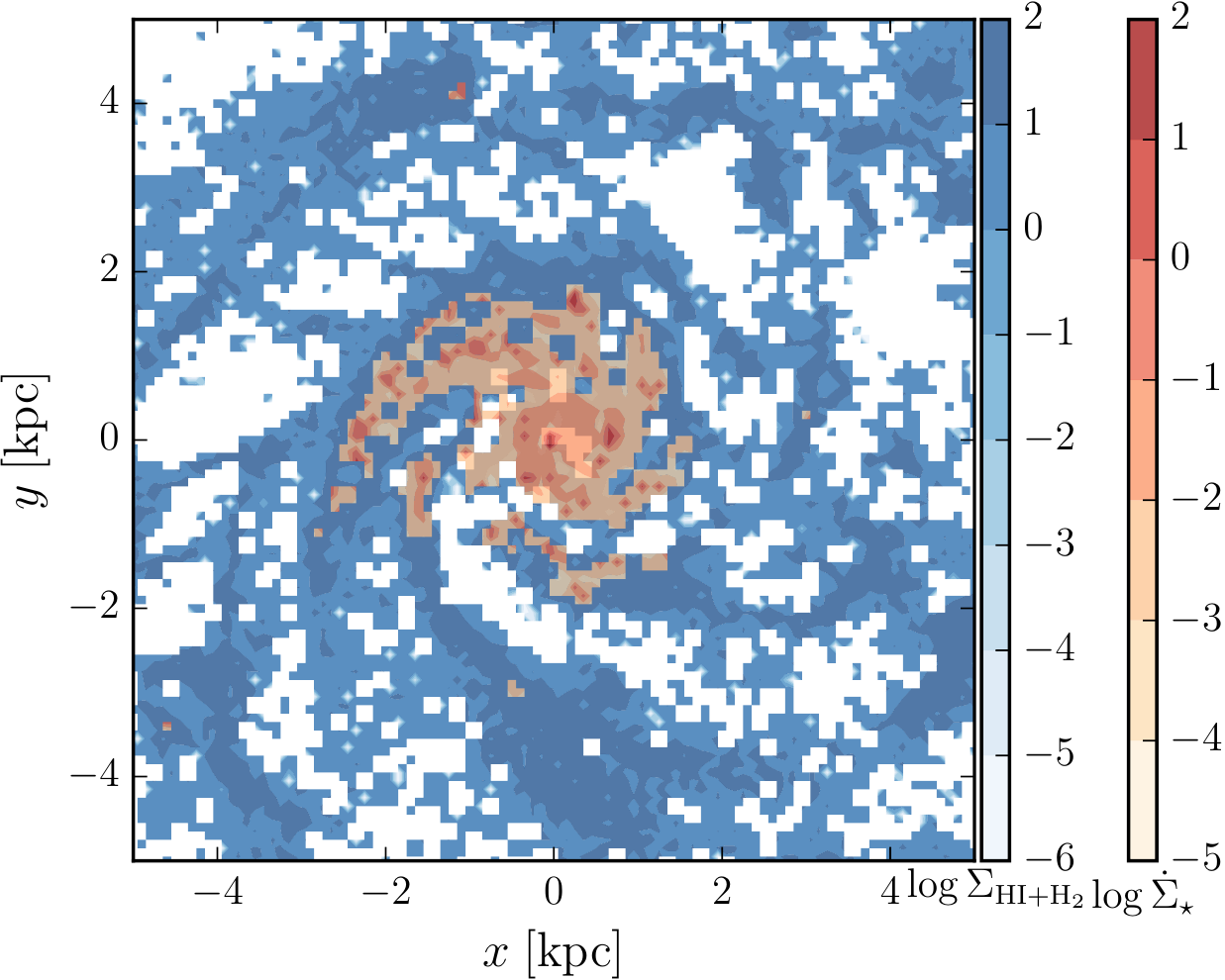}
	\caption{Example of one of our maps, made from a Milky-Way mass simulated galaxy at $z \approx 0$ \citep[galaxy {\bf m12i} from][]{Hopkins2014}, with 100 pc pixels.  Neutral hydrogen surface density, $\Sigma_{\rm HI + H_2}$ [${\rm M_{\odot}}$ pc$^{-2}$] and instantaneous gas star formation rate $\dot\Sigma_\star$ [${\rm M_{\odot}}$ yr$^{-1}$ kpc$^{-2}$] are colored in blues and reds, respectively.  Spiral arms and increasing density towards the galactic core are clearly visible, and the instantaneous star formation rate is seen to closely trace the densest gas structures.}
	\label{mapfig}
\end{figure}

It has also been suggested that the KS relation may evolve with redshift or have a metallicity dependence \citep{Schaye2004, Bouche2007, Papadopoulos2010, Dib2011, Gnedin2011, Scoville2016}.  These are not entirely independent quantities, as metallicity generally increases as galaxies process their gas through generations of stars over cosmic timescales. Because the presence of metals results in more efficient gas cooling, and can thus aid in the transition from diffuse ionized and atomic species to dense molecular gas \citep{Hollenbach1999}, \citet{Schaye2004} and \citet{Krumholz2009b} have argued that there is a metallicity-dependent gas surface density cutoff below which the KS relation becomes steeper.  \citet{Krumholz2009b} attribute the dependence to the gas column needed to self-shield molecular gas for a given metallicity.  As well gas metallicity has been argued to weakly modulate the specific strength of stellar feedback, as SNe couple slightly less momentum into their immediate stellar surroundings since more of their energy is able to radiate away quickly \citep{Cioffi1988, Martizzi2015, Richings2016}.  \citet{Scoville2016} found evidence of shorter depletion timescales for molecular gas at higher redshifts for galaxies both on and above the `star formation main sequence', perhaps due to the rapid accretion required to replenish the gas reservoirs. 

Large-volume cosmological simulations often use the KS law as a sub-grid prescription for star formation, both because of the prohibitive computational complexity of including all of the physics relevant on the scales of star-forming regions, and their inability to resolve even the most-massive giant molecular clouds $\sim 10^6$  M$_\odot$ \citep[e.g.][]{Mihos1994, Springel2003}.  Even idealized disk simulations that have resolution on the order of 100 pc, but are unable to resolve a multiphase ISM, employ star formation prescriptions that assume low star formation efficiencies \emph{a priori} or implement KS laws indirectly \citep{Li2006, Wada2007, Schaye2008, Richings2016}.  It has been shown that assuming a power-law star formation relation on the resolution scale can imprint a power-law relation of identical slope on the galactic scale \citep{Gnedin2014}, demonstrating the importance of employing physically motivated sub-grid star formation prescriptions that produce kpc-scale relations with the `correct' slope if the relevant physical processes cannot be treated directly.  With advances in computing power, and the ability to execute increasingly complex simulations with more physics at higher mass resolution, cosmological simulations have only recently been able to \emph{predict} the KS relation generically as a result of the physics incorporated in the simulations at the scales of molecular clouds \citep[e.g.][]{Hopkins2011,Hopkins2013b,Hopkins2014,Agertz2015}.  

Including realistic feedback physics in simulations that resolve GMC scales is critical to understanding the KS relation due to the multitude of competing physical effects spanning a wide range of scales.  \rev{While some simulations have argued that the KS relation can be obtained without explicit feedback \citep[e.g.][]{Li2005, Li2006, Wada2007}, these generally depend on either (a) transient and short-lived initial conditions (e.g. simulations starting from strong initial turbulence or a smooth disk, where once turbulence decays and fragmentation runs away, some additional source of ``driving" or ``GMC disruption" must be invoked), or (b) suppressing runaway fragmentation numerically (e.g. ``by hand'' setting very slow star formation efficiencies at the grid scale, or inserting explicit sub-grid models for star formation calibrated to the KS relation on GMC or galaxy scales, or adopting artificial/numerical pressure or temperature floors or fixed gravitational softening in the gas that prevent densities from increasing arbitrarily). Many of these authors do acknowledge that ``feedback'' is likely necessary to provide either the initial conditions or grid-scale terms in their simulations, even if not explicitly included (similarly, see e.g. \citealt{Robertson2008, Colin2010, Kuhlen2012, Kraljic2014}). Indeed, a large number of subsequent, higher-resolution numerical experiments (on scales ranging from kpc-scale ``boxes'' to cosmological simulations) which run for multiple dynamical times and allow fragmentation to proceed without limit have consistently shown that absent feedback, the galaxy-scale KS law has a factor $\sim 100$ higher normalization than observed \citep[see e.g.][]{Hopkins2011, Kim2011, Ostriker2011, Shetty2012, Kim2013, Kim2015, Dobbs2015, Benincasa2016, Forbes2016, Hu2017, Iffrig2017}.}

In this paper, we explore the properties and emergence of the KS relation in the FIRE\footnote{http://fire.northwestern.edu} (Feedback In Realistic Environments) simulations \citep{Hopkins2014}.  Specifically, by producing mock observational maps of the spatially-resolved KS law, we investigate the form of the relation when considering several different tracers of the star formation rate and gas surface densities, and we characterize its dependence on redshift, metallicity, and pixel size. The FIRE simulations are well suited for understanding the physical drivers of the KS relation as they sample a variety of galactic environments and a large dynamic range in physical quantities (chiefly, gas and star formation rate surface densities), and they directly (albeit approximately) incorporate stellar feedback processes that may be crucial for the emergence, and maintenance of the KS relation over cosmological timescales.  In the past, they have been used to investigate the effects of various microphysics prescriptions on galaxy evolution, the formation of giant gas clumps at high redshift, and the formation of galaxy discs, among other topics \citep{Oklopcic2017,KungYi2016,Ma2017}.

%------------------------------------------------------------------------------------
\section[]{Simulations \& Analysis Methods}\label{meth}

In the present analysis, we investigate the star formation properties of the FIRE galaxy simulations originally presented in \citet{Hopkins2014}, \citet{Chan2015}, and \citet{Feldmann2016}, which used the Lagrangian gravity + hydrodynamics solver {\sc gizmo} \citep{Hopkins2013a} in its pressure-energy smoothed particle hydrodynamics (P-SPH) mode \citep{Hopkins2013a}.  All of the simulations employ a standard flat $\Lambda$CDM cosmology with $h \approx$ 0.7, $\Omega_{\rm M} = 1 - \Omega_\Lambda \approx 0.27$, and $\Omega_{\rm b} \approx 0.046$.  The galaxies in the simulations analyzed in this paper range in $z \approx 0$ halo masses from $9.5 \times 10^9$ to $1.4 \times 10^{13}$ M$_\odot$, and minimum baryonic particles masses $m_b$ of $2.6 \times 10^2$ to $3.7 \times 10^5$ M$_\odot$.  For all of the simulations, the mass resolution is scaled with the total mass such that the characteristic turbulent Jeans mass is resolved.  As well, the force softening is fully adaptive, scaling with the particle density and mass as 
\begin{equation}
\delta h \approx 1.6 \; {\rm pc} \left( \frac{n}{\rm cm^{-3}}\right)^{-1/3} \left( \frac{m}{10^3  \, \rm M_\odot}\right)^{1/3} ,
\end{equation}
where $n$ is the number density of the particles, and $m$ is the particle mass.  Consequently, the simulations are able to resolve a multiphase ISM, allowing for meaningful ISM feedback physics.  This is crucial because the vast majority of star formation occurs in the most massive GMCs due to the shape of the GMC mass function \citep{Williams1997}.

The stellar feedback physics implemented in these simulations include approximate treatments of multiple channels of stellar feedback: radiation pressure on dust grains, supernovae (SNe), stellar winds, and photoheating; a detailed description of the stellar feedback model can be found in \citet{Hopkins2014}.  Star particles in the simulations each represent individual stellar populations, with known ages, metallicities, and masses.  Their spectral energy distributions, supernovae rates, stellar wind mechanical luminosities, metal yields, etc. are calculated directly as a function of time using the {\sc starburst99} \citep{Leitherer1999} stellar population synthesis models, assuming a \citet{kroupa2002} IMF.

In these simulations, the galaxy- and kpc-scale star formation efficiencies are {\em not} set `by hand'.  Star formation is restricted to dense, molecular, self-gravitating regions according to several criteria:
\begin{itemize}  
\item The gas density must be above a critical threshold, $n_{\rm crit} \sim 50 $ cm$^{-3}$ in most runs \citep[and 5 cm$^{-3}$ in those from][]{Feldmann2016}.  

\item The molecular fraction $f_{\rm H_2}$ is calculated as a function of the local column density and metallicity using the prescription of \citet{Krumholz2011}, and the molecular gas density is used to calculate the instantaneous SFR (see below).

\item Finally, we identify self-gravitating regions using a sink particle criterion at the resolution scale, specifically requiring $\alpha \equiv \delta v^2\delta h/Gm_{\rm gas}(<\delta r) < 1$ on the smallest resolved scale around each gas particle ($\delta h$ being the force softening or smoothing length).  
\end{itemize}
Regions that satisfy all of the above criteria are assumed to have an instantaneous star formation rate of 
\begin{equation}\label{eq:sfr}
\dot{\rho}_* = \rho_{\rm mol}/t_{\rm ff},
\end{equation} 
i.e. 100 percent efficiency per free-fall time.  As a large fraction of the dense ($n > n_{\rm crit}$), molecular ($f_{\rm H_2} \sim 1$) gas is not gravitationally bound ($\alpha  > 1$) at any given time, the \emph{global} star formation efficiency $\epsilon$ is less than 100 per cent ($\epsilon < 1$) despite the assumed \emph{local, instantaneous} star formation efficiency per free-fall time being 100 per cent.  We will show below that the KS relation, with its much lower \emph{global, time-averaged} star formation efficiency ($\epsilon \lesssim 0.1$), emerges as a result of stellar feedback preventing dense gas from quickly becoming self-bound and forming stars and disrupting gravitationally bound star-forming clumps on a timescale less than the local free-fall time. We stress here that the emergent KS relation is \emph{not} a consequence of the star formation prescription employed in the simulations.

In Appendix \ref{sec:appendix:baryonic.physics} we demonstrate this explicitly.  We ran several tests restarting one of the standard FIRE simulations with varying physics and star formation prescriptions.  For any reasonable set of physics, only variation in the strength of the feedback affected the galactic star formation rates, because the simulated galaxies self-regulate their star formation rates via feedback.  A number of independent studies have also shown that once feedback is treated explicitly, the predicted KS law becomes independent of the resolution-scale star formation criterion \citep[][]{Saitoh2008, Federrath2012, Hopkins2012b, Hopkins2013d, Hopkins2013b, Hopkins2013c, Hopkins2016, Agertz2013}.

To quantify the spatially resolved KS relation in the simulations, we analyze data from snapshots spanning redshifts  $z = 0-6$.  The standard FIRE snapshots from \citealt{Hopkins2014} and the dwarf runs in \citet{Chan2015} are roughly equipartitioned amongst redshift bins $z \sim 3 - 6$, $2.5 - 1.5$, $1.5 - 0.5$, and $< 0.5$, whereas the snapshots of halos from \citet{Feldmann2016} have redshifts evenly spaced between $2 < z < 6$ (these were run to only $z \sim 2$).   To compare the snapshots with observational constraints of the KS relation, we made star formation rate and gas surface density maps of each snapshot's central galaxy.  We summed the angular momentum vectors of the star particles in the main halo of each snapshot to determine the galaxy rotation axis and projected along this axis to generate face-on galaxy maps.  The projected maps were then binned into square pixels of varying size, ranging from 100 pc to 5 kpc on a side.  Only particles within 20 kpc above or below the galaxy along the line of sight were included in the maps (this captures all of the star-forming gas, but excludes distant galaxies projected by chance along the same line-of-sight in the cosmological box).  An example of the resulting maps can be found in Figure~\ref{mapfig}, which shows maps of the neutral gas surface density and the instantaneous star formation rate surface density in the {\bf m12i} simulation from \citet{Hopkins2014}, at redshift $z \approx 0$ with 100 pc pixels.

Using the star particle ages, we calculated star formation rates averaged over the previous 10 and 100 Myr, correcting for mass loss from stellar winds and other evolutionary effects as predicted by {\sc starburst99} \citep{Leitherer1999}.  We also considered the instantaneous star formation rate of the gas particles (defined above). A time-averaging interval of 10 Myr was chosen because this approximately corresponds to the timescale traced by recombination lines such as $\rm H\alpha$, whereas UV and FIR emission traces star formation over roughly the past 100 Myr \citep[e.g.][]{Kennicutt2012}.\footnote{Directly computing SFR indicators from the simulations \citep[e.g.][]{Hayward2014,Sparre2015} rather than computing the SFR averaged over the past 10 or 100 Myr would provide a more accurate comparison with the observed KS relation, but doing so would considerably expand the scope of this work, so we leave it to a future study.}  The instantaneous star formation rate of the gas particles covers a larger range of star formation rates because it is not constrained at the low end directly by the mass resolution of our simulations; it is a continuous quantity intrinsic to the gas particles themselves, which is sampled at each time-step to determine if the gas particles form stars.  This quantity best demonstrates the direct consequences of feedback on the gas {\em in situ} by locally tracing the star formation rate, whereas the other two SFR tracers are more analogous to observables.  
%%%%%%%%%%% FIGURE 2 - VARIOUS FORMULATIONS OF KS LAW PANEL
\begin{figure*}
	\centering
	\includegraphics[width=0.99\textwidth]{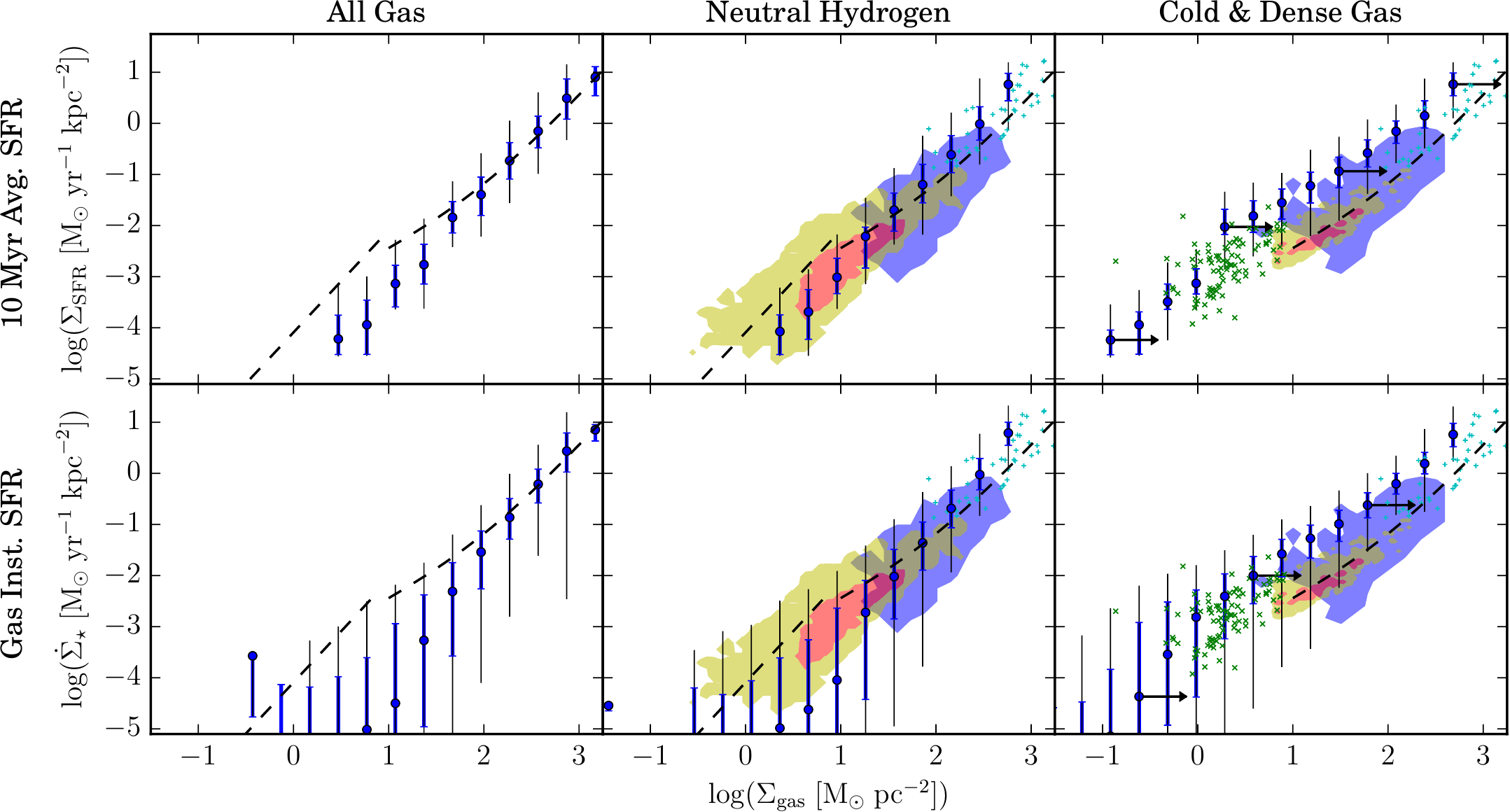}
	\caption{KS relation in the FIRE runs in 1 kpc$^2$ pixels, binned by $\Sigma_{\rm gas}$, for several gas and star formation ``tracers''.  Neutral hydrogen is $\sim \Sigma_{\rm HI + H_2}$, and Cold \& Dense gas includes particles with T $< 300$ K {\em and} $n_{\rm H} > 10$ cm$^{-3}$ ($\sim \Sigma_{\rm H_2}$).  The gas instantaneous star formation rate is calculated directly from the gas particles in each pixel, whereas the 10 Myr average star formation rate is calculated from the young star particles in each pixel.  \rev{Median} values of the pixel distribution are plotted in bins of gas surface density, with thick (thin) error bars denoting the 25-75\% (5-95\%) range of resolved star formation in the bins.  {In both the middle and right columns, the} yellow and red shaded regions denote observational data from  \citet{Bigiel2008}, and the blue shaded region illustrates the region spanned by the resolved observations from \citet{Kennicutt2007}. Observations of high-redshift dusty star-forming galaxies from \citet{Genzel2010} (cyan +'s) and molecular gas data from \citet{Verley2010} (green x's, right column only) are also included.  \rev{All observations have been re-calibrated with the \citet{Narayanan2012} variable $X_{\rm CO}$ interpolation function, as described in Section \ref{obsref}.}  The star formation relations derived in Sections \ref{disc:high} and \ref{disc:low} are plotted with dashed black lines, using the fiducial values assumed there and $\Sigma_\star = 10^2$ M$_\odot$ pc$^{-2}$. \rev{The simulations' kpc-scale total gas and neutral hydrogen KS relations are consistent with observational constraints within the uncertainties. The simulation relation computed using the Cold \& Dense tracer is systematically offset from the observed molecular gas relation; this may be due to the Cold \& Dense tracer underestimating the molecular fraction (see text and Appendix \ref{sec:appendix:molefrac}). The lower limit error bars in the Cold \& Dense gas panels indicate the 0.5~dex uncertainty in our conservative estimator of molecular gas.}}
	\label{tracers}
\end{figure*}

The gas surface density tracers were also chosen on the basis of observable analogues, including all gas, neutral hydrogen gas (total H${\rm I}$ + H$_2$ column, accounting for metallicity and He corrected), and ``Cold \& Dense'' gas which we specifically define here and throughout this paper as gas with $T < 300$ K and $n_{\rm H} > 10$ cm$^{-3}$.  These roughly correspond to the total gas (including the ionized component), atomic + molecular gas (${ \rm HI + H_2}$), and cold molecular gas reservoirs observed in galaxies.  We have opted to use the aforementioned approximation for the molecular gas component rather than reconstruct the $f_{\rm H_2}$ predicted by the \citet{Krumholz2011} model (which is not output in the snapshots) as the $f_{\rm H_2}$ model assumes a simplified geometry at the resolution scale, that can get the local optical depth quite wrong\footnote{For the purposes of our star formation criteria, however, this is not an issue for the vast majority of cases.  Due to the steepness of the exponential attenuation of the local UV field, we care only whether, strictly speaking, the gas is optically thin or thick, but the exact value of $\tau$ is not especially important, as any optical depth $\tau \gg 1$ effectively yields $\exp{(-\tau)} \ll 1$, and $\tau \ll 1$ similarly yields $\exp{(-\tau)} \approx 1$.}.  \rev{We explore the differences between the Cold \& Dense gas tracer and the \citet{Krumholz2009b} $f_{\rm H_2}$ (which was the basis for \citealt{Krumholz2011}) in a small number of snapshots, as well as with other empirical estimators such as those adopted in \citet{Leroy2008}, in Appendix~\ref{sec:appendix:molefrac}.}  A more detailed analysis of the true molecular fraction of the gas would require a careful radiative transfer post-processing, which we leave to a later work. Furthermore, as {\sc GIZMO} lacks a detailed implementation of any chemical network, important to determining low temperature cooling, and instead uses approximate cooling tables, we may get the temperature wrong by a factor of a few below $\sim 1000$ K (this error should have no \emph{dynamical} effect in the simulations as this cool gas already effectively has no pressure compared to the bulk of the gas at higher, more reliable temperatures).  Past work by \citet{Richings2016} has indeed shown that metallicity and radiation field on large scales have far larger effects on star formation rates than including detailed low-temperature chemical networks.  

We acknowledge that because of the rather strict density and temperature criteria, the lack of any additional considerations, e.g. to the local UV field or the geometry of the gas, \rev{and our ``low" star formation gas density threshold of 50~cm$^{-3}$, we appear to} underestimate the molecular gas column \rev{as measured by the Cold \& Dense gas tracer (and other estimators calculated at the particle scale) by} up to a factor of a dex, which is incidentally on the order of the uncertainty in the observational CO to H$_2$ conversion factor $X_{\rm CO}$ \citep{Bolatto2013}.  This likely results in a corresponding underestimation of the local gas depletion time and overestimation of star formation efficiency.  In Appendix~\ref{sec:appendix:molefrac}, we show explicitly that the ``molecular'' fraction based on the ``Cold \& Dense'' criteria is significantly less than the molecular gas fraction computed using two other relations for $f_{\rm H_2}$ versus total neutral gas surface density: that from \citet{Leroy2008} \rev{(which is based on \citealt{Blitz2006})} and the relation from \citet{Krumholz2009b} applied at the kpc-scale for total gas surface densities above their atomic-to-molecular transition thresholds. Notably, applying \citet{Krumholz2009b}, with updates from \citealt{Krumholz2011}, at the particle scale produces a similar underestimation of the molecular gas column of $\sim$0.5~dex like the Cold \& Dense gas tracer.  The difficulty of estimating local (at the particle scale) column depths for shielding likely contributes to the discrepancy for both the Cold \& Dense gas tracer and \citet{Krumholz2009b} fit applied at the few-pc scale.

\rev{These empirical fits for $f_{\rm H_2}$, based in part on the stellar surface densities and scale heights and gas metallicity, suggest that the FIRE simulations are producing correct star formation rates for large-scale properties of the ISM, e.g., mid-plane pressure, implying that the discrepancy in the Cold \& Dense gas tracer lies with the dense end of the gas phase structure at the particle scale, and not with kpc-scale properties of the galaxies.}  However, we believe that the scaling relations based on the Cold \& Dense' tracer are robust, since this discrepancy results in a \emph{consistent} bias in the normalization of `cold' gas.  \rev{Again, a more accurate calculation would involve radiative transfer post-processing including a chemical network, which would allow us to directly predict the molecular hydrogen fraction and $X_{\rm CO}$, which we intend to pursue in future work.}

Other quantities are calculated as the mass-weighted average in each pixel, including the gas metallicity\footnote{In this paper, we take solar metallicity to be $Z_{\odot} \approx 0.0142$ when scaling metallicities \citep{Asplund2009}.} $Z$, the Keplerian velocity $v_c$, and the dynamical angular velocity $\Omega$, defined here as
\begin{equation}
\Omega = \frac{v_c}{R} = \frac{(GM(<R))^{1/2}}{R^{3/2}},
\end{equation}
 where $R$ is the galactocentric radius of the pixel and $M(<R)$ is the total mass enclosed within that radius (and $G$ is the gravitational constant). These quantities allow us to investigate the dependence of star formation on gas phase metallicity, approximate the optical depth of star-forming regions, and relate galactic dynamical times to star-forming regions.

In our analysis we treat pixels from all simulations and all times equally, unless otherwise stated.  However we wish to examine only ensembles of pixels with well-resolved SFR distributions.  Recalling that each of our simulations has a fixed baryonic particle mass, $m_{b}$, we discard pixels which contain $<$3 gas particles; for a pixel size $l$, this means only gas surface densities $\Sigma_{\rm gas} > 3\times10^{-3}\,{\rm M_\odot}\,{\rm pc^{-2}}\,(m_{b}/1000\,{\rm M_\odot})\,(l/{\rm kpc})^{-2}$ will be considered. However, in the example above ($m_{b}\sim1000\,{\rm M_\odot}$, $l\sim$\,kpc), the observed \citet{Kennicutt1998} relation gives a typical star formation surface density $\Sigma_{\rm SFR} \sim 10^{-7}\,{\rm M_\odot}\,{\rm yr}^{-1}\,{\rm kpc}^{-2}$ at this minimum $\Sigma_{\rm gas}$, so in $\sim 10\,$Myr, the expected number of $m_{b}\sim1000\,{\rm M_\odot}$ star particles formed will be just $0.001$. Obviously, the distribution of star formation rates will not, then, be resolved (even if the simulations capture the mean star formation rate correctly, the discrete nature of star formation means only 1 in 1000 pixels will have a star particle, while 999 have none).  Thus, to ensure that the pixels we examine from each simulation at a given gas surface density have a well-resolved SFR distribution, we adopt the following criteria: (1) we first calculate the {\em mean} $\Sigma_{\rm SFR}$ per pixel from each simulation, for all their pixels with a given number of gas particles (fixed $\Sigma_{\rm gas}$); (2) we estimate the average number $\langle N_{\star}(\Delta t) \rangle$ of star particles this would produce in the time $\Delta t$ ($10$ or $100$\,Myr, as appropriate); (3) if this is $<1 (=N_{\rm min, \star})$, we discard all pixels which contain this number or fewer gas particles. For the example above, for $\Delta t=10\,$Myr ($100\,$Myr), this requires $>500$ ($>50$) gas particles per pixel for a ``resolved'' star formation rate. We have repeated this exercise using instead the observed KS relation (instead of the predicted one), to estimate the resolved thresholds, and find it gives nearly identical results. We have also verified that changing the threshold $N_{\rm min, \star}$ by an order of magnitude in either direction does not change any of our conclusions here; we note too that the {\em average} star formation rates from low-resolution simulations continue to agree well with our higher-resolution simulations down to $\langle N_{\star}(\Delta t) \rangle$ as low as $\sim0.001$.

\rev{We believe it important to reassert that {\bf pixels with no star formation contribute to all of the plotted points in our KS relation}.  We are discarding sets of pixels (those with and without star formation) that do not have at least one young star particle on average at a given gas surface density, to ensure that all of our plotted data points are drawn from well-resolved distributions of star formation (including zero star formation) at a given gas surface density.}

We are careful that this prescription does not introduce bias into our star formation distributions at a given gas surface density.  If we were to consider the distribution of depletion time ($\Sigma_{\rm gas}/\dot\Sigma_\star$) \rev{across all gas surface densities}, this method \emph{would} bias us towards shorter depletion time \rev{by discarding all the pixels below the gas surface density that definitely produces at least one new star particle in the past 10 (or 100) Myr}.  However, we are investigating the distribution of SFRs in bins of gas surface density \rev{for ensembles of pixels from a number of individual galaxy simulations}.  To do so, we examine many snapshots from \rev{each individual} simulation and consider whether the SFR \rev{distribution is well sampled by} the ensemble of pixels from all of those snapshots at a given gas surface density.  \rev{If at that gas surface density in the whole ensemble of pixels from that single simulation, there are at least $N_\star$ (we have chosen one here\footnote{\rev{We have confirmed that this approach does not bias the average SFR surface density values by repeating the analysis requiring only an average of $N_\star = 0.001$ star particles per pixel. In this case, however, the distribution of SFR surface density at a given gas surface density is poorly sampled because of Poisson noise.}}) new star particles produced \emph{on average}, then we believe we are able to say something meaningful about the distribution of star formation rates in that bin of gas surface density for that ensemble of pixels.}  In combining only the sets of pixels from individual simulations with resolved SFR distributions at a given gas surface density, we thus avoid biasing our aggregated SFR (and by extension, depletion time) \emph{distributions} in each bin of gas surface density.
\subsection{Observational Data}\label{obsref}
 Comparing with observations, we compiled resolved KS observations from a large number of papers at various resolution scales commensurate with our mock observational maps. For our 1~kpc ``fiducial'' scale maps of the KS relation, we compare our neutral gas surface density results with a combination data from \citet{Kennicutt2007}, \citet{ Bigiel2008}, \citet{ Genzel2010}; we compare our 1~kpc ``Cold \& Dense'' gas surface density results with the appropriate $\rm H_2$ results from these studies, as well as those from \citet{Verley2010}.  For exploring the effects of pixel size, we also used these molecular gas data to compare with our 500-pc maps, as these observations had varying resolution scales ranging from 500~pc to slightly larger than 1~kpc.  For our galaxy-averaged 5~kpc maps, we used data from \citet{Kennicutt1998}, \citet{ Kennicutt2007}, \citet{ Genzel2010}, \citet{ Shapiro2010}, \citet{ Wei2010}, \citet{ Freundlich2013}, \citet{ Tacconi2013} and \citet{Amorin2016} .  Finally, for our highest-resolution investigations at 100~pc, we compared with high-resolution observations from \citet{Blanc2009}, and \citet{ Onodera2010}. For exploring the star formation efficiency in this work, in the form of the Elmegreen-Silk relation, we compare our 1~kpc-scale maps with observations from \citet{Kennicutt1998} and \citet{Daddi2010}.  
 
 No distinction is made between the many estimators of SFR used in the aforementioned papers; they are simply taken at face value.  \rev{However, we re-calibrate $X_{\rm CO}$ in the observationally-inferred $\Sigma_{\rm H_2}$ data points across all the aforementioned resolved KS studies with an interpolation function taken from \citet{Narayanan2012}, of the form $X_{\rm CO} = {\rm min}[4, 6.75 \times W_{\rm CO}^{-0.32}]\times 10^{20}$~cm$^{-2}$/(K km s$^{-1}$), independent of metallicity\footnote{\rev{Though their full interpolation function included a metallicity dependence, we assume solar metallicity for simplicity.}}.  To correct the quoted $\Sigma_{\rm H{\scriptsize I} +H_2}$ measurements, we decomposed the total column into atomic and molecular components (the latter then being corrected in the manner of the $\Sigma_{\rm H_2}$'s) using data in the references themselves, where available, and assuming a molecular fraction fit from \citet{Leroy2008} where necessary.  We explore the effects of variations of the assumed $X_{\rm CO}$ on the \rev{(dis)agreement} with our simulations in Appendix~\ref{sec:appendix:xco}, finding $\sim 0.5$~dex uncertainty due to the uncertainty in $X_{\rm CO}$.  In the case of the Elmegreen-Silk relation observations from \citet{Kennicutt1998} and \citet{Daddi2010}, being unable to separate out the dynamical times, we recalibrate \citet{Kennicutt1998} only to a constant $X_{\rm CO} = 2 \times 10^{20}$~cm$^{-2}$/(K km s$^{-1}$), consistent with \citet{Bigiel2008}.  Data from \citet{Daddi2010} have not been altered due to the extensive efforts made therein to calibrate $X_{\rm CO}$ across their dataset.}

 %------------------------------------------------------------------------------------
\section{KS Relation in the Simulations}\label{results}

\subsection{Dependence of the KS Relation on Star Formation and Gas Tracers}\label{main-result}

Figure \ref{tracers} demonstrates \rev{how a KS-like power-law relation self-consistently emerges (recall that the assumed instantaneous star formation efficiency of dense, gravitationally bound `molecular' gas is 100 per cent per local free-fall time) in the FIRE simulations irrespective of specific choice of star formation or gas tracer}.  Two of our star formation rate tracers, the 10 Myr-averaged and gas instantaneous star formation rates, yield very similar KS relations.  The points denote the \rev{median value} of the star formation rate distribution \rev{in that gas surface density bin}.  The thick (thin) error bars in Figure \ref{tracers} denote the 25-75\% (5-95\%) inclusion interval in the distribution of the star formation rates of pixels in that bin of gas surface density, effectively the $\pm 1\sigma$ $(\pm 2\sigma)$ scatter.  The $1\sigma$ scatter of our 10 Myr-averaged SFR, neutral gas, KS relation is $\sim$0.4 dex, in line with quoted scatters from \citet{Bigiel2008} and \citet{Leroy2013}.

More restrictive gas tracers (e.g. taking gas with $T < 300$ K and $n_H > 10 $ cm$^{-3}$, instead of all atomic + molecular gas) yield shallower power-law slopes.  This is intuitive because by placing more restrictions on the gas column, we are taking pixels at a given star formation rate and moving them to lower gas surface densities (to the left) by reducing what gas contributes to the overall gas column density.  The restrictions are non-linear: at high surface densities, the gas is predominately molecular, and added restrictions do little to change the participating gas column, whereas at low surface densities, relatively little of the gas column may remain after making these additional cuts.  Little difference is seen between the star formation distributions in $\Sigma_{\rm SFR}-\Sigma_{\rm gas}$ space when considering the surface densities of all gas (including the ionized component) versus neutral gas (first and second columns of Figure \ref{tracers}) because the contribution of ionized gas to the total gas column is small in regions where significant star formation is occurring. In contrast, there is a marked change in slope of the KS relation when moving from the surface density of neutral hydrogen gas to that of Cold \& Dense gas ($T < 300$ K, $n_H > 10$ cm$^{-3}$), with the slope shifting from $\sim 1.7$ to $\sim 1.2$ for the gas instantaneous star formation rate.  This is due to the fact that significant amounts of star formation can occur in ``small'' pockets of molecular gas, relative to the overall gas column, yielding a shallower slope than when considering neutral gas.

\rev{The neutral gas surface density} KS relation in the FIRE simulations is consistent with the corresponding spatially resolved observational data, as represented by the shaded regions and points in the panels of Figure~\ref{tracers}.  No observational range has been included for ``All Gas'' observations as this is not typically observed; nevertheless, our data suggest that little change would be evident, as again, ionized gas does not usually contribute significantly to the column of star forming gas.  \rev{There is significant, though consistent, disagreement} between the simulations and observations for our Cold \& Dense gas surface density because our ``Cold \& Dense'' \rev{appears to underestimate the expected molecular gas surface density by $\sim 0.5-1$ dex.  In Appendix~\ref{sec:appendix:molefrac}, we explore the uncertainty in the molecular gas mass estimate by comparing the Cold \& Dense gas tracer with other empirical estimators for the molecular fraction of our pixels, such as the dependence on the mid-plane gas pressure used in \citet{Leroy2008}, adapted from earlier work \citep{Blitz2006}, and the self-shielding-based method from \citet{Krumholz2009b}.  There, we see that the Cold \& Dense gas tracer (which is calculated on a per-particle basis) appears to under-predict molecular gas fractions by $\sim 0.5-0.7$~dex across surface densities of 1-100 M$_\odot$ pc$^{-2}$ compared to the kpc-averaged empirical estimators.  This suggests that although our star formation rates are appropriate given the large-scale properties of the ISM (e.g., mid-plane pressure and dust opacity), we are under-predicting the mass of gas at the highest densities, either by converting it into stars too quickly as it crosses our star formation density threshold or by incorrectly approximating the cooling and shielding properties of the densest gas.  However, as this under-prediction appears to be \emph{consistent} across the range of gas surface densities explored, we believe the \emph{form} of the KS relation to be robust and have added arrows indicating this $\sim 0.5$~dex underestimate whenever results based on the Cold \& Dense gas tracer are presented.  Exploring this further is beyond the scope of this work and is the subject of a forthcoming study forward modeling dense gas tracers in FIRE-2.}

Interestingly, the distribution of star formation rates in the FIRE simulations overlaps with that of damped Ly$\alpha$ systems (DLAs) observed at high redshift by \citet{Rafelski2016}.  Though we appear to see analogues to these systems at 1 kpc$^2$ pixel sizes, we leave it to a future work to investigate the detailed physical properties of these systems.

\subsubsection{Elmegreen-Silk Relation (Alternative KS Law)}\label{subsec:ES}
%%%%%%%%%%% FIGURE 3 - VARIOUS FORMULATIONS OF EM-SILK LAW PANEL
\begin{figure*}
	\centering
	\includegraphics[width=0.99\textwidth]{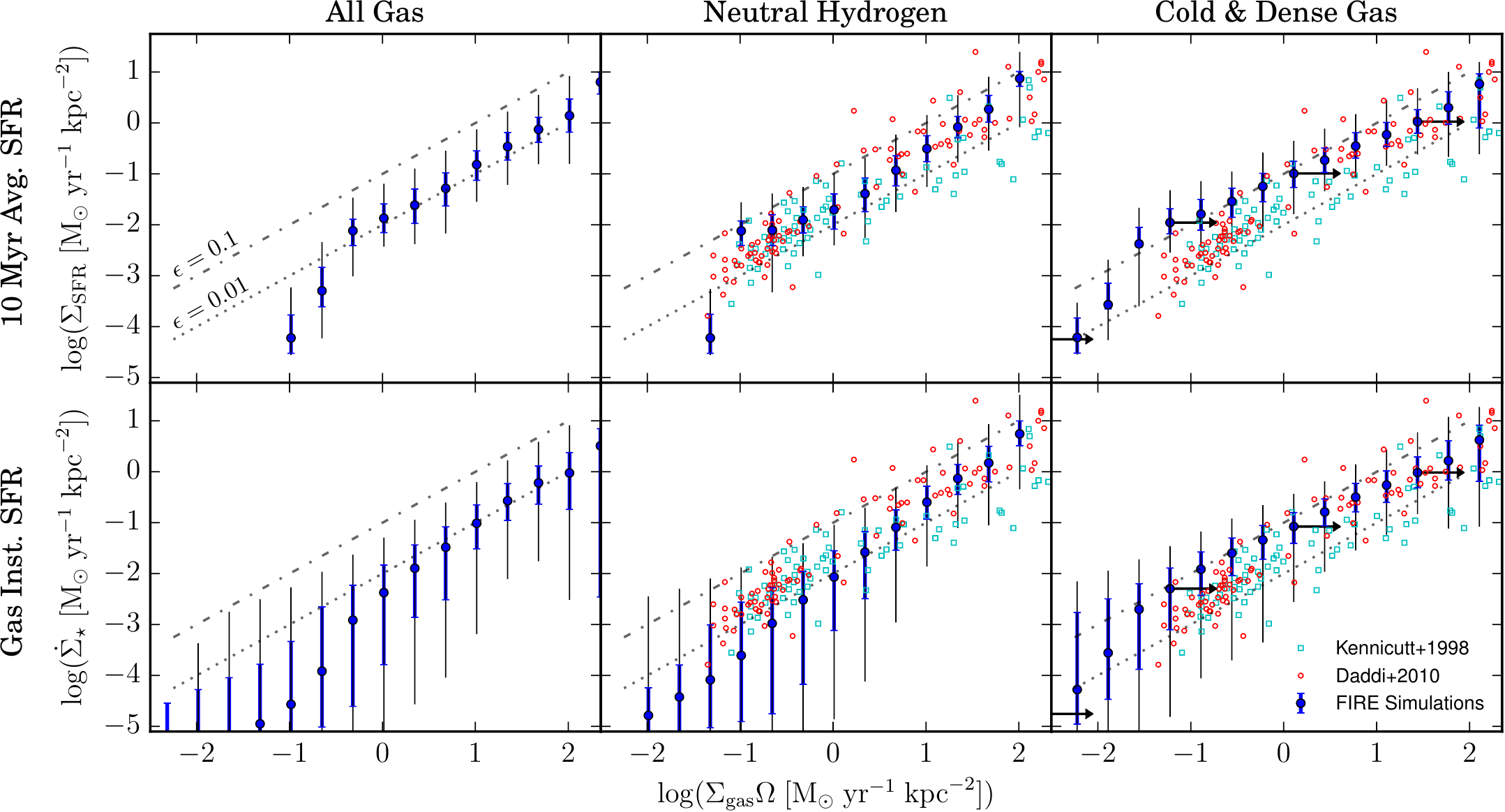}
	\caption{Elmegreen-Silk relation in the FIRE runs in 1 kpc$^2$ pixels, in the style of Figure \ref{tracers}. Lines of constant star forming efficiency are plotted, with $\epsilon = (0.01, 0.1)$.  Unfilled cyan squares and red circles are observational data from \citet{Kennicutt1998} and \citet{Daddi2010}, respectively. \rev{The data from \citet{Kennicutt1998} have been recalibrated to a $X_{\rm CO}$ value consistent with \citet{Bigiel2008}, but those of \citet{Daddi2010} are unaltered.} The simulated galaxies have kpc-scale star formation efficiencies increasing from $\sim 1\%$ (all gas) to $\sim 10\%$ (Cold \& Dense gas) as denser gas tracers are selected.} %\cch{Add observational points to legend.}}
	\label{tracers_eff}
\end{figure*}

Alternatively, in Figure~\ref{tracers_eff}, we probe the global efficiency of gas turning into stars in a dynamical time according to
\begin{equation}\label{ES_rel}
\dot\Sigma_\star = \epsilon \Sigma_{\rm gas} \Omega ,
\end{equation}
where $\epsilon$ represents the ``star formation efficiency'' on kpc scales; this relation is known as the Elmegreen-Silk relation \citep{Elmegreen1997, Silk1997}.  We see systematic agreement with the neutral hydrogen gas surface density Elmegreen-Silk relation in the FIRE simulations compared to observations where $\epsilon$ ranges between $10^{-3} - 1$.  We find kpc-averaged star formation efficiencies of $\epsilon \sim 0.01 - 0.1$ consistently for our entire range of star formation rate surface density.  Dashed black lines indicate constant efficiencies between 0.01 and 1.  Without feedback, one would expect to see $\epsilon \sim 1$.  

Our efficiencies for the molecular gas formulation of the Elmegreen-Silk relation are likely over-estimated by as much as 1 dex, at efficiencies between a few and a few tens of percent, because of our systematic underestimation of the mass of ``Cold \& Dense'' gas \rev{(see also the discussion at the end of Section \ref{main-result} and in Appendix~\ref{sec:appendix:molefrac})}.  However, since this is likely consistent across gas surface densities, we believe that the relative constancy of global star formation efficiency $\epsilon$ across $\Sigma_{\rm gas}$ is robust.  \rev{Error bars of 0.5~dex (which are likely conservative) indicate this underestimation in the Cold \& Dense gas panels.}  Even so, we find \rev{consistency} at the high end of the observed efficiencies using the molecular gas formation of the Elmegreen-Silk relation, where our molecular fraction finally converges to near unity.

\subsubsection{100 Myr Averaged Star Formation Rate}\label{subsec:100myr}
%%%%%%%%%%% FIGURE 4 - KS LAW PANEL FOR 100 MYR AVG SFR
\begin{figure}
	\centering
	\includegraphics[width=0.45\textwidth]{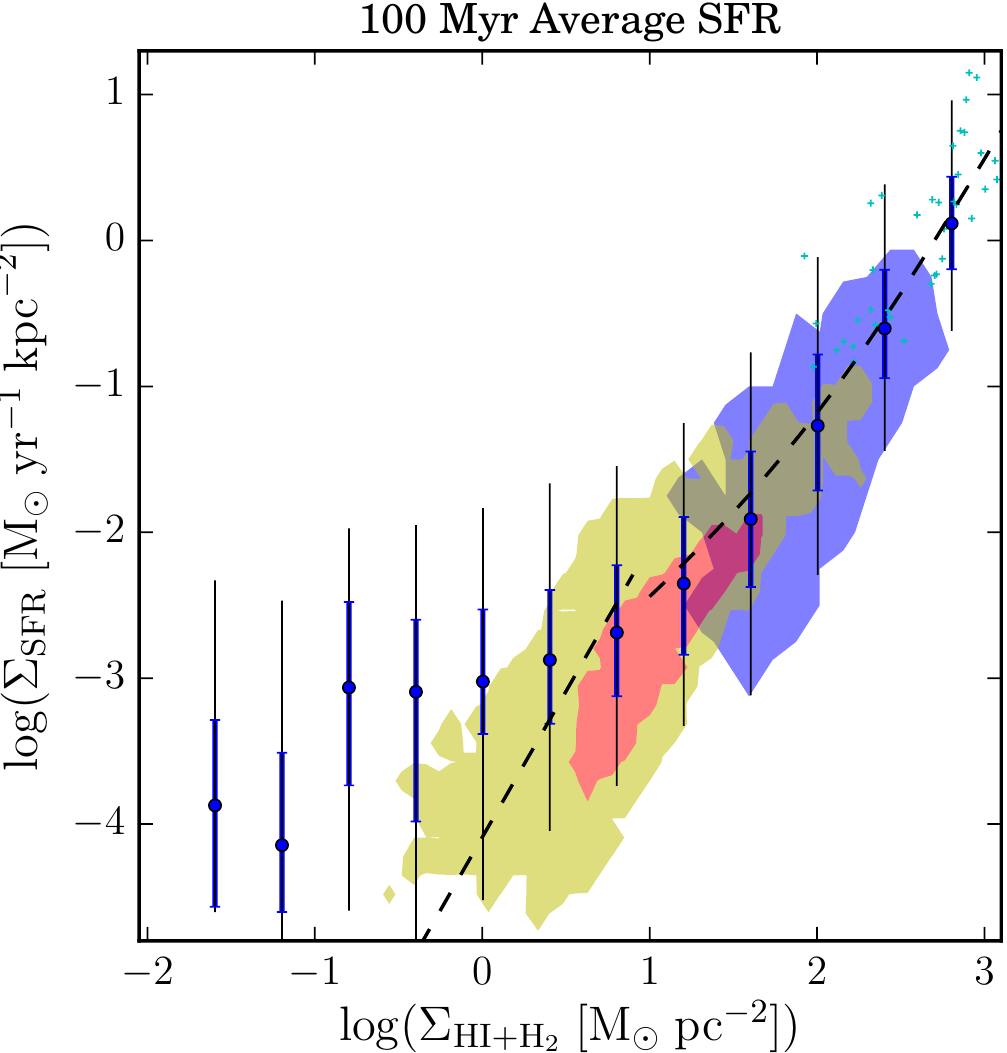}
	\caption{KS relation in the FIRE simulations for the 100 Myr-averaged star formation rate in 1 kpc$^2$ pixels, as Figure \ref{tracers}.  The {observational data shaded regions and points for the neutral (\rev{atomic + molecular}) gas} are those {from \citet{Kennicutt2007}, \citet{Bigiel2008}, and \citet{Genzel2010} as described in Figure \ref{tracers},} measured with $\sim 10$ Myr tracers.  At high $\Sigma_{\rm gas}$, the $\sim 100$ Myr {average SFRs} agree well  {with the $\sim 10$ Myr observations (and by extension our $\sim 10$ Myr-averaged SFRs)}.  At low $\Sigma_{\rm gas}$, the $\Sigma_{\rm SFR}$ from the $\sim 100$ Myr tracer flattens.  This appears to stem from a breakdown in the correlation between 100 Myr-old stars and the observed gas tracers, either from migration or other dynamical effects (e.g. mergers or strong outflow events).}
	\label{100Myr_fig}
\end{figure}
%%%%%%%%

In Figure \ref{100Myr_fig}, we see a clear flattening of the 100 Myr-averaged star formation rate surface density relative to the 10 Myr average, for neutral hydrogen columns at low gas surface densities, $\Sigma_{\rm gas} \lesssim 1$ M$_\odot$ pc$^{-2}$.  This is ascribable to effects discussed in Sections \ref{sec-res} and \ref{sec-met}, where individual or small numbers of young star particles are scattered into regions of very low gas surface density that are not actually forming stars.  Moreover, dynamical changes in star-forming regions over the averaging period (100 Myr) cause gas complexes to dissipate and produce small numbers of star particles left in now-diffuse galactic environments.  At high gas surface densities, $\Sigma_{\rm gas} > 10$ M$_\odot$ pc$^{-2}$, the 100 Myr average star formation rate surface densities agree well with the shorter time-scale estimators.

\subsection{Pixel Size Dependence}\label{sec-res}
%%%%%%%%%%% FIGURE 5 - RESOLUTION DEPENDENCE
\begin{figure*}
	\centering
	 %DIFDELCMD < \includegraphics[width=0.99\textwidth]{figs/KSlaw_10myrcolddense_resdep_new.pdf}
%DIFDELCMD < 	%%%
  \includegraphics[width=0.99\textwidth]{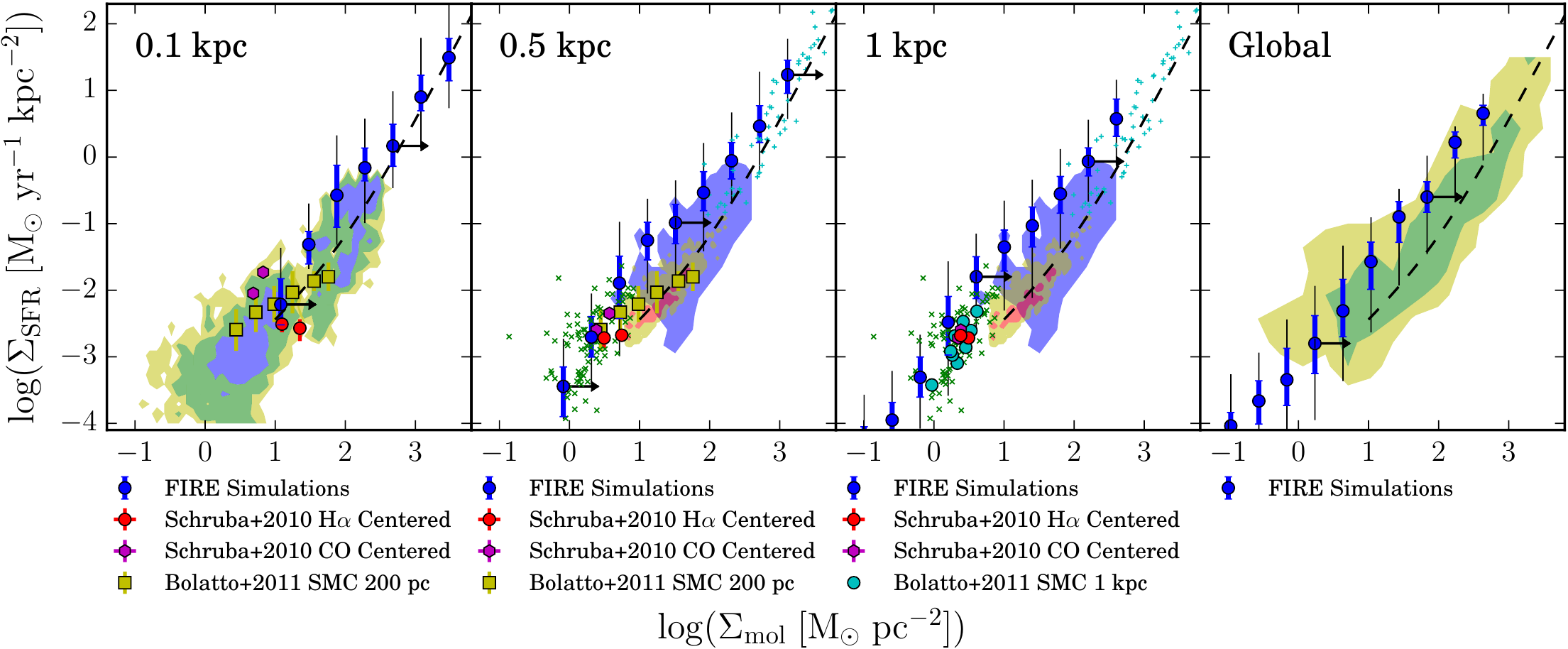}
	 \caption{Pixel size dependence of the molecular KS relation in the FIRE simulations, compared with selected observations. $\Sigma_{\rm mol}$ is the surface density of Cold \& Dense gas ($T < 300$ K, $n_{\rm H} > 10$ cm$^{-3}$); $\Sigma_{\rm SFR}$ is the 10 Myr averaged star formation rate surface density.  Plotted points and error bars are as in Figure \ref{tracers}. {Shaded} regions {and points (small green x's and cyan +'s)} for both 500~pc and 1~kpc {denote $\rm H_2$ observations}, as {described in Figure \ref{tracers}, because} a number of the source observations lie between those two resolutions. {In the 0.1~kpc panel, shaded regions (yellow, green, blue) denote the individual inclusion contours (100, 90, 50\%)} from {\citet{Blanc2009} and \citet{Onodera2010}. For the global KS relation panel}, the {yellow (green) shaded areas are} the {70\% (50\%) inclusion regions for global molecular-KS observations compiled from the references listed in Section \ref{obsref}. To calculate our global} KS relation, {we sum the SFR and Cold \& Dense gas mass values} in {the map} and {divide the sums by $\pi R^2_{1/2 \star}$, where $R_{1/2 \star}$ is the stellar half-mass radius.  \rev{All observations have been re-calibrated with the \citet{Narayanan2012} variable $X_{\rm CO}$ interpolation function, as described in Section \ref{obsref}.  Lower limit error bars indicate our $\sim 0.5$~dex uncertainty in molecular gas mass using the conservative Cold \& Dense gas estimator.}  Various points explicitly enumerated below the panels correspond to} observations {of M33} from  \citet{Schruba2010} {and the SMC from \citet{Bolatto2011}, testing the} scale dependence {of} the {KS relation in those environments}. The star formation relations derived in Section \ref{disc:high} are plotted with dashed black lines {as in Figure~\ref{tracers}} .  For each pixel size, there is a threshold in $\Sigma_{\rm mol}$ below which star formation is poorly resolved given even our highest mass resolution, as described in Section \ref{meth}.  Above this threshold, the KS relation {in the simulated galaxies} exhibits no systematic trend with pixel size despite dynamical processes that might be expected to break down the correlations between young stars and gas on small scales (prominent at small pixels sizes).}
	\label{res-dep}
\end{figure*}
%%%%%%%%%%% FIGURE 6 - EM-SILK REDSHIFT DEPENDENCE
\begin{figure}
	\centering
	\includegraphics[width=0.45\textwidth]{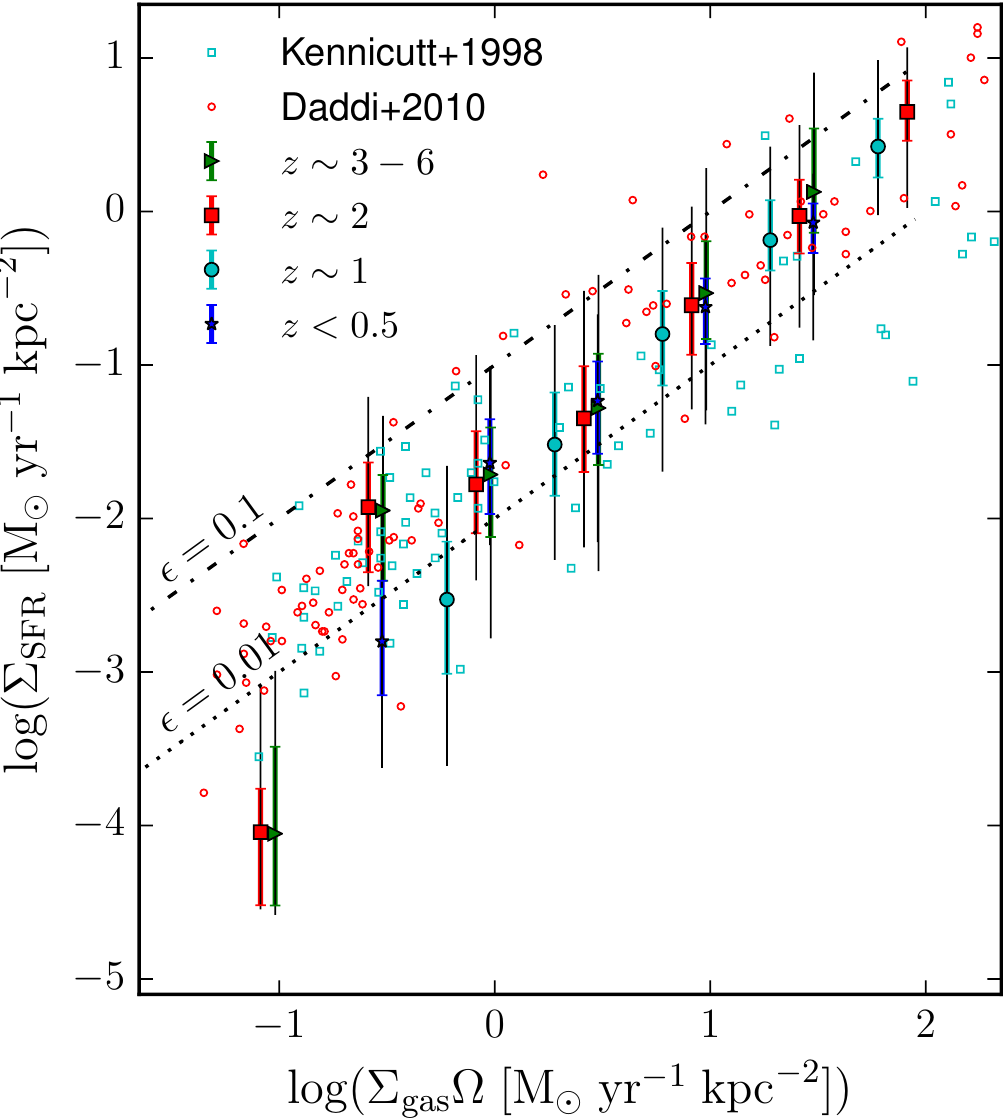}
	\caption{Elmegreen-Silk relation binned in redshift at 1 kpc$^2$ pixel size.  \rev{Median} values of the 10 Myr average $\Sigma_{\rm SFR}$ are plotted in bins of $\Sigma_{\rm HI + H_2} \Omega$, in the style of Figure \ref{tracers}. {Observations from \citet{Kennicutt1998} (unfilled cyan squares)} and  \citet{Daddi2010} {(unfilled red circles), in addition to} dotted lines representing constant star formation efficiencies $\epsilon$, {are included}.  No significant dependence on redshift is seen: the range of data in each bin is greater than any systematic difference between bins.}
	\label{redeff}
\end{figure}
%%%%%%%%%%% FIGURE 7 - METALLICITY AND REDSHIFT DEPENDENCE
\begin{figure}
	\centering
	\includegraphics[width=0.45\textwidth]{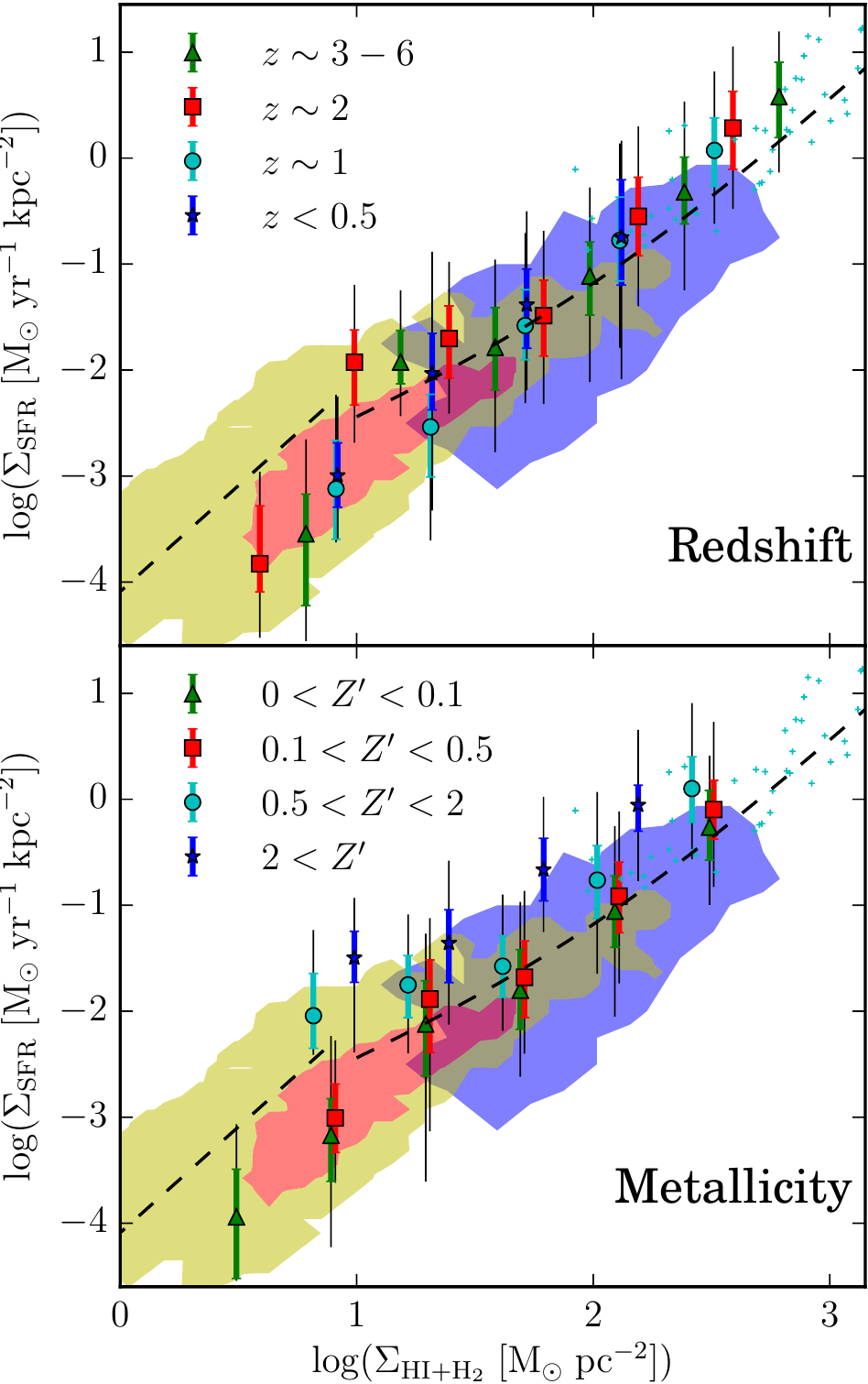}
	\caption{KS relation binned in redshift and metallicity at 1 kpc$^2$ pixel size.  In both panels, within their respective redshift and metallicity bins, \rev{median} values of the 10 Myr average $\Sigma_{\rm SFR}$ are plotted in bins of $\Sigma_{\rm HI + H_2}$, in the style of Figure \ref{tracers}. {Shaded regions and small cyan triangles denote representative} observations {from \citet{Kennicutt2007}, \citet{Bigiel2008}, and \citet{Genzel2010}}, and dotted lines represent the derived star formation relations, {all} as {described} in Figure \ref{tracers}. {\bf Top Panel:} Simulation snapshots binned by redshift, with markers denoting different epochs.  No significant dependence on redshift is seen -- the range of data in each bin is greater than any systematic difference between them.  {\bf Bottom Panel:} Pixels from snapshots binned by gas metallicity, with markers indicating intervals in $Z' = Z_{\rm gas}/Z_{\odot}$. {A} weak {positive correlation between} $\Sigma_{\rm SFR}$ {and metallicity} is seen at all $\Sigma_{\rm HI + H_2}$, {though the dependence is weak compared to the scatter in each gas bin.}  No metallicity-dependent cutoff is evident.}
	\label{rednmetals}
\end{figure}
%%%%%%%%%%% FIGURE 8 - PARAMETERIZED METALLICITY DEPENDENCE
\begin{figure}
	\centering
	 %DIFDELCMD < \includegraphics[width=0.45\textwidth]{figs/KSlaw_1kpc_ZSFRexpect_neut_n1.pdf}
%DIFDELCMD < 	%%%
  \includegraphics[width=0.47\textwidth]{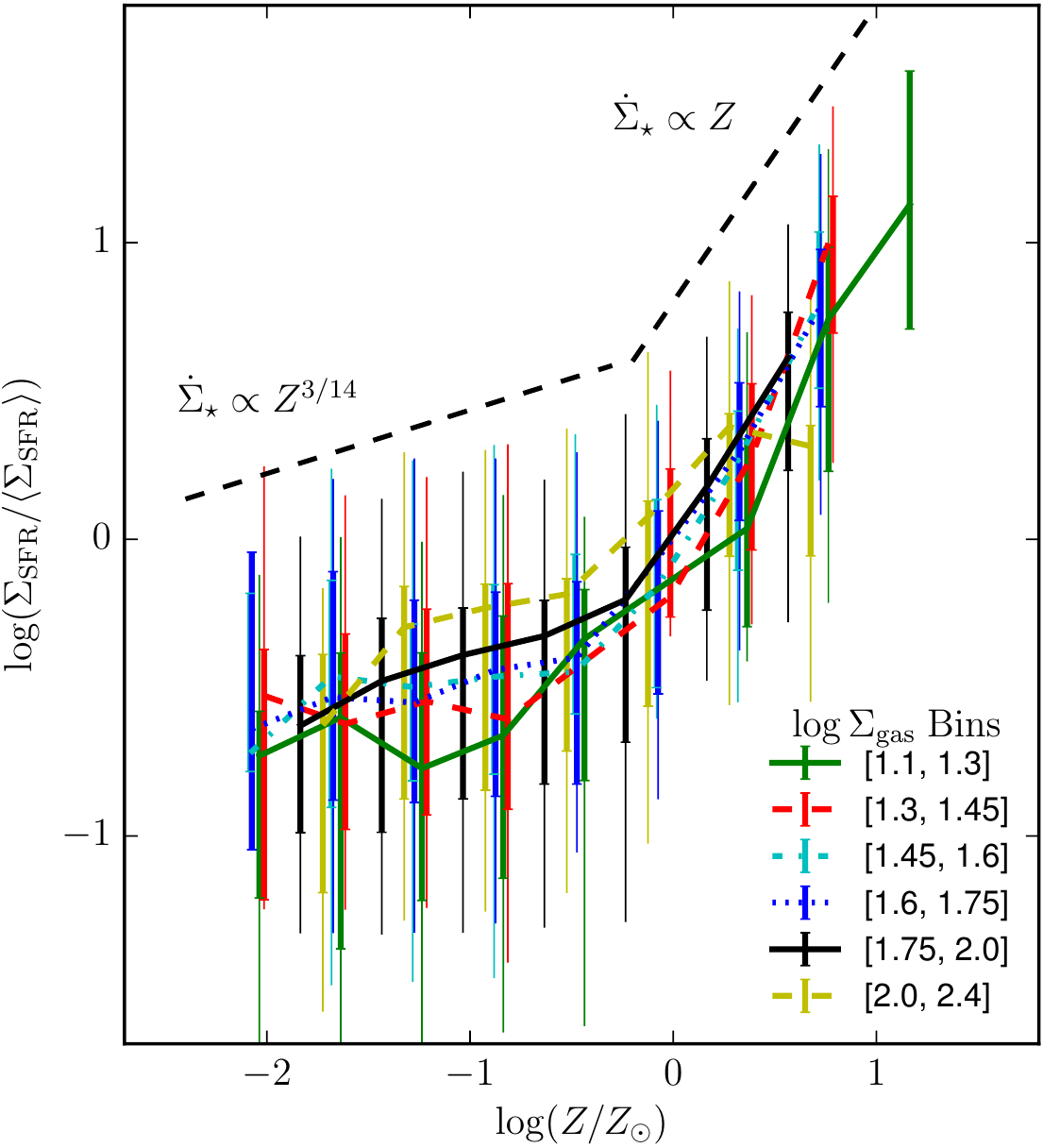}
	 \caption{Star formation rate dependence on metallicity, binned by $\Sigma_{\rm HI + H_2}$ at 1 kpc$^2$ pixel size.  Points are \rev{median} values of the 10 Myr-average $\Sigma_{\rm SFR}$ normalized by the average $\Sigma_{\rm SFR}$ {for a given bin} in {$Z/Z_\odot$ in} each $\Sigma_{\rm HI + H_2}$ bin. {Thick (thin) error} bars denote the {25-75\% (5-95\%)} range for resolved star formation in each bin.  A weak dependence on metallicity is seen for all gas surface densities {for sub-solar metallicities}, as demonstrated by the dashed black line of slope {$\Sigma_{\rm SFR} \propto Z^{3/14}$} . {But a stronger, nearly linear, dependence is seen for all gas surface densities above solar metallicity values, evidenced by the dashed line with slope $\Sigma_{\rm SFR} \propto Z$.}}
	\label{metal_dep}
\end{figure}
%DIF > %%%%%%%%%% FIGURE 9 - PARAMETERIZED DYNAMICAL TIME DEPENDENCE
 \begin{figure}
	\centering
	\includegraphics[width=0.47\textwidth]{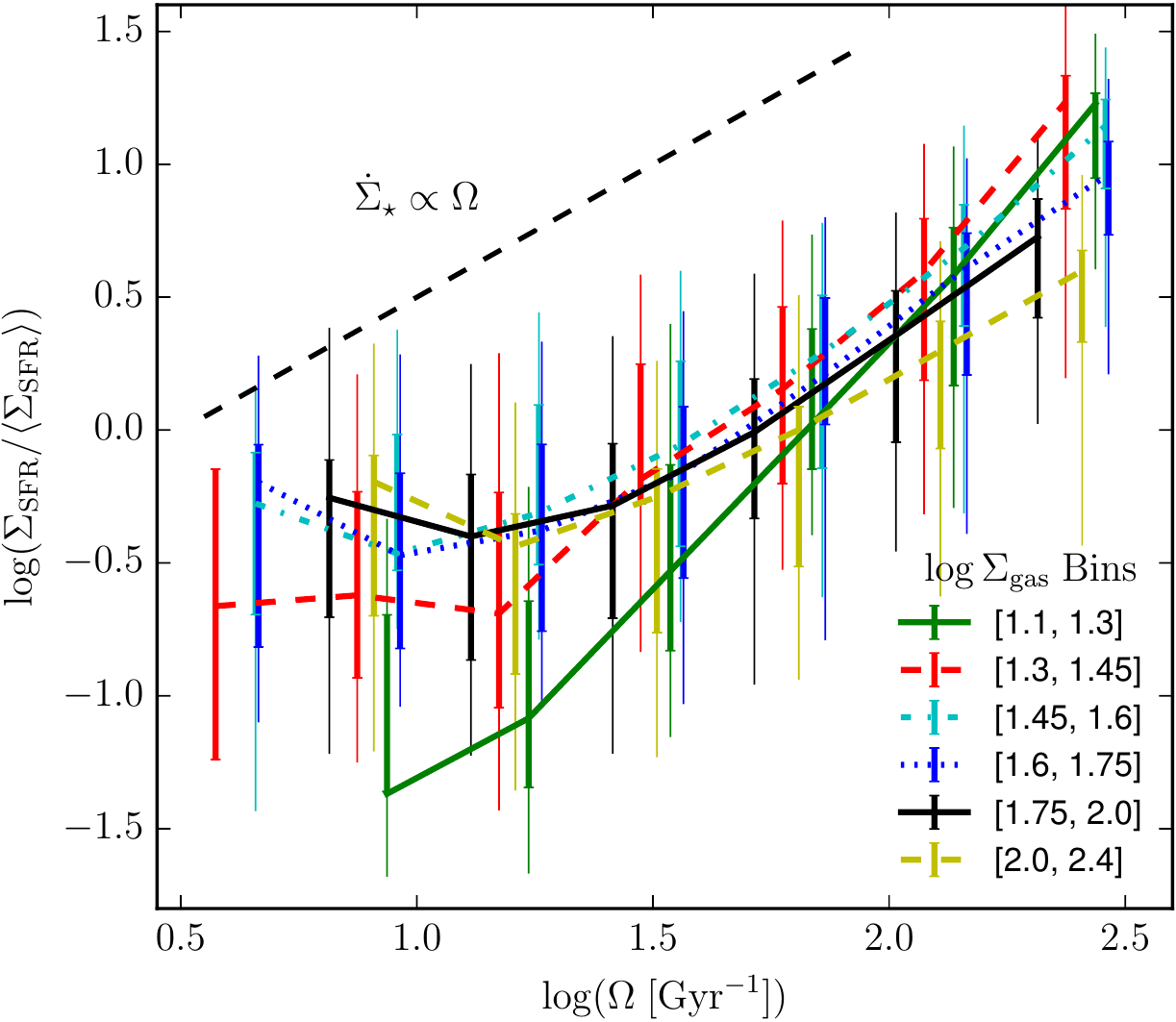}
	\caption{{SFR surface density versus $\Omega$ (1/dynamical time), binned by $\Sigma_{\rm HI + H_2}$ at 1 kpc$^2$ pixel size.  Points are median values of the 10 Myr-average $\Sigma_{\rm SFR}$ normalized by the average $\Sigma_{\rm SFR}$ for a given bin in $\Omega$ in each $\Sigma_{\rm HI + H_2}$ bin.  Thick (thin) error bars denote the 25-75\% (5-95\%) range for resolved star formation in each bin.  A strong inverse dependence on dynamical time is seen for all gas surface densities despite considerable scatter, as demonstrated by the dashed black line with slope $\Sigma_{\rm SFR} \propto \Omega = 1/t_{\rm dyn}$.}}
	\label{omega_dep}
\end{figure}

The KS relation that we find in the FIRE simulations does not appear to have a significant dependence on pixel size (i.e. map resolution) for pixels with sufficiently resolved gas and star formation rate tracers ( $\gtrsim$ few gas/star particles per pixel), as shown in Figure \ref{res-dep}.  Over the range of pixel sizes we investigate, 100 pc - 5 kpc (0.01 - 25 kpc$^2$), the slope of the power law varies only weakly between $\sim 1$ and $\sim 4/3$.  At the low end of the relation in $\Sigma_{\rm gas}$, we expect the scatter to grow as Poisson statistics become important when only a few star particles are present in the pixels on average.  However, because we exclude poorly-sampled pixels, this simply manifests as a lower limit to the plotted $\Sigma_{\rm gas}$ for smaller pixels sizes (see Section \ref{meth}).  

\rev{In terms of slope, our simulated relations agree with the observed relations for the various pixel sizes considered, but again, the simulated and observed relations are systematically offset, likely because the ``Cold \& Dense'' gas tracer systematically underestimates the column density of molecular gas (by $\sim 0.5-1$ dex) relative to that computed using the fits of \citet{Leroy2008} and \citet{Krumholz2009b}, as already discussed above.  In addition to the shaded regions shown in previous plots,} we also compare directly with the results of \citet{Schruba2010} and \citet{Bolatto2011}.  \citet{Schruba2010} compare the KS relation found for varying aperture scales in M33, centered either on $\rm H\alpha$ or CO peaks.\footnote{\citet{Schruba2010} do not tile M33 with their apertures, but this does not appear to matter except at their smallest aperture scales.}  Their results vary weakly with pixel size, except at their smallest aperture scale $\sim 75$ pc.  Similarly, \citet{Bolatto2011} observed the KS relation in the Small Magellanic Cloud (SMC), and averaged their results with 200~pc and 1~kpc apertures to investigate its dependence on averaging scale.  Their results are also consistent with our simulations, considering that the ``Cold \& Dense'' tracer underestimates the molecular fraction by as much as a dex for surface densities above 10 M$_\odot$ pc$^{-2}$.  We see, however, a slightly steeper KS relation at pixel sizes of 100 and 500~pc (their data at 200~pc lie between these scales) and a slightly shallower relation at kpc scales than \citet{Bolatto2011}.

To compare with the global KS relation observed by a number of studies (see Section~\ref{obsref} for references), we summed the total 10~Myr star formation rate and Cold \& Dense gas mass in each map and then divided these sums by the area circumscribed by the stellar half-mass radius calculated for each snapshot in order to produce analogous global KS results.  Our global molecular KS relation is nearly identical in form to that observed, but like other results involving our Cold \& Dense gas tracer, our gas surface densities are appear to be underestimated by \rev{$\sim 0.5-0.7$~dex} for $\Sigma_{\rm mol} \gtrsim 1 \; {\rm M}_\odot \; {\rm pc}^2$.

At our smallest pixel size (100 pc), {however}, none of our simulations are able to adequately sample star formation at gas surface densities  $\lesssim 10$ M$_\odot$ pc$^{-2}$, {given our mass resolution;} this regime is where observations exhibit the largest scatter.  At least three processes cause the correlation of the star forming gas and young star particles to break down on scales less than $l \sim 500$ pc.  (1) The relative velocities between star forming gas and the young stars they produce cause them to wander into different pixels, thus they become uncorrelated on a pixel-by-pixel basis, when $v_p \sim l / \Delta t$.  For 100 pc pixels and {10 } Myr time bins, this is a relative velocity of only {$\sim 10$ } km s$^{-1}$ ({1} km s$^{-1}$ for {$\Delta t \sim 100$ Myr} ), so we would expect significant scatter to arise from this effect at the smallest pixel sizes.  (2) Dynamical processes affecting gas and star particles, like dispersion of GMCs, major mergers, or SNe, over the time bin (i.e. 10 --100 Myr) cause greater fluctuations from the power law average as pixel size decreases.  (3) When considering small ($< 1$ kpc) pixels, scatter is caused by the stochastic nature of the star formation in the simulations. {Above $\sim 10$ M$_\odot$ pc$^{-2}$, the simulated and observed relations again agree \rev{in terms of slope}, but the normalization is offset by $\sim 1$ dex due to the Cold \& Dense gas tracer underestimating the molecular gas fraction.
} 

\subsection{Redshift Independence}

We find no significant redshift dependence of either the KS or Elmegreen-Silk relations in the FIRE galaxies. The insensitivity to redshift in the simulations can be seen in Figure~\ref{redeff} and the top panel of Figure \ref{rednmetals}, where the snapshots are colored by redshift bin ($z < 0.5$, $0.5 - 1.5$, $1.5 - 2.5$, $3 - 6$) and the 10 Myr-averaged $\Sigma_{\rm SFR}$ and neutral gas surface density are considered.  \rev{Similarly, no redshift dependence was seen for the Cold \& Dense gas version of either relation; consequently, and due to the extensively discussed issues with the Cold \& Dense tracer, these results are not shown.}  Some scatter is seen in the average values between redshift bins, but any dependence on redshift is much smaller than the range of the data itself. The absence of any redshift dependence persists for all measures of star formation rate.  Though the absolute amount of star formation varies with redshift, the correlation between gas column and star formation rate surface density, and star formation efficiency, remains consistent.  

\subsection{Metallicity Dependence}\label{sec-met}
We see in the bottom panel of Figure \ref{rednmetals} evidence of a weak dependence on metallicity for the KS relation in the FIRE runs.  For all neutral gas surface densities, more metal-rich gas exhibits elevated star formation rates, with an admittedly large scatter (there is significant overlap in $\Sigma_{\rm SFR}$ range for various $Z'$ bins).  \rev{At low gas surface densities ($\sim 1-10$~M$_\odot$ pc$^{-2}$), the strength of the metallicity dependence appears to be consistent with the predictions of \citet{Krumholz2009b} and \citet{Dib2011}.}  Interestingly, none of our forms of the KS law exhibit a notable metallicity-dependent cutoff in star formation, as some models predict \citep{Krumholz2009b}.  For our 10 and 100 Myr-averaged star formation rates, this may be an issue of adequately sampling star formation rates in the ``cutoff'' regime of $\dot\Sigma_\star \sim 10^{-(3-4)}$ M$_\odot$ yr$^{-1}$ kpc$^{-2}$.  However, for the well-resolved instantaneous star formation rate, the form of the star formation relation does not change at all for any of the metallicity bins in the ``cutoff'' regime that \citet{Krumholz2009b} find.  The instantaneous star formation rate tracer, as with the averaged star formation rate tracers, presents higher star formation rates for metal-enriched gas even at these low gas surface densities, but the change is smooth, rather than a ``threshold'' effect.  \rev{The metallicity dependence does not appear to be strongly dependent on gas surface density, and star formation rates remain consistently positively correlated with metallicity above 10 M$_\odot$ pc$^{-2}$, differing from the model of \citet{Dib2011}, which argues for a negative correlation owing to the metallicity dependence of pre-supernova feedback (e.g. momentum coupling in winds)}.

Figure~\ref{metal_dep} illustrates the strength of the metallicity dependence.  Binning pixels by gas surface density $\Sigma_{\rm gas}$ and normalizing by the average star formation rate in each $\Sigma_{\rm gas}$ bin, we find that star formation rate surface density increases weakly with metallicity {below approximately solar metallicity and considerably stronger above solar metallicity} across all $\Sigma_{\rm gas}$ bins.  This presentation of the data normalizes out the $\Sigma_{\rm gas}$ dependence to highlight the much weaker $Z$ dependence.  A by-eye fit of a power law with {$\dot\Sigma_\star \propto Z'^{3/14}$ for sub-solar metallicities and $\propto Z$ above solar metallicity} is plotted as a dashed black line. {In the sub-solar regime, this} slope is much shallower than the slope derived later in Section \ref{disc:low} but on the order of the predicted metallicity dependence of SNe feedback's momentum injection \citep[ranging from $\sim 1/10 - 3/14$;][]{Cioffi1988, Martizzi2015}\footnote{As described in \citet{Hopkins2014}, when SNe explode in regions such that their cooling radii will be unresolved (common in some of the lower-resolution simulations here with particle masses $\gtrsim 10^4$ M$_\odot$), the ejecta are assigned a terminal momentum based on the detailed individual explosion models from \citet{Cioffi1988}, which scale as $p_t \propto Z^{3/14}$.  However, the metallicity dependence in Figure \ref{metal_dep} persists if we restrict only to our highest-resolution simulations.}.  \rev{A lack of a strong dependence on gas surface density appears to indicate that the metallicity dependence of star formation due to pre-supernova feedback effects are subdominant compared to that of supernova feedback in the FIRE simulations \citep{Dib2017}}.  Above approximately solar metallicity, a stronger, nearly linear dependence appears.  This dependence is more consistent with that derived in Section~\ref{disc:low}, but the reasons for its appearance are unclear and warrant future investigation. Though the scatter within bins is quite large, {the weak (and stronger)} dependence are rather robust across all gas surface densities.  \rev{A similar dependence on metallicity was found in the Cold \& Dense gas version of Figures~\ref{rednmetals} \& \ref{metal_dep}, and for brevity, we do not include them.}

\subsection{Dependence on Dynamical Time}\label{omega-dep}

 {In a similar manner to Figure~\ref{metal_dep}, we investigate the dependence of SFR surface density on $\Omega (= 1/t_{\rm dyn})$ in Figure~\ref{omega_dep}.  Again normalizing the 10 Myr SFR surface density to the average SFR surface density within bins of $\Sigma_{\rm gas}$, we see a strong nearly linear dependence of SFR surface density on $\Omega$, as expected both for a turbulently supported ISM, as discussed in Section~\ref{disc:high} (see Eq.~\ref{rawturb}), and the thermally supported regime discussed in Section~\ref{disc:low} (see Eq.~\ref{rawtherm}).  Interestingly this persists for all gas surface densities, connecting the low- and high-gas-surface-density regimes.  The dependence on $\Omega$ appears to be weaker at higher gas surface densities, which may point to an increasing prevalence of ``turbulent'' Toomre stability (see Eq.~\ref{SF_eqn_high}, with no explicit $\Omega$ dependence).}

 %------------------------------------------------------------------------------------
\section{Physical Interpretation}\label{disc}
On the scales of tens or hundreds of millions of years, it is possible to understand star formation as an equilibrium process (on galactic scales) in which the inputs of either momentum injection from stellar feedback (at high gas surface density) or energy from photoheating (at low gas surface density) balance gravitational collapse.

\subsection{High Gas Surface Density Regime}\label{disc:high}

In our analysis, there is a marked transition in the star formation rate distribution at gas column densities above $\Sigma_{\rm gas} \sim 100$ M$_\odot$ pc$^{-2}$.  Above this threshold, almost all the gas forms stars on or very near the KS power law.  Here, supernova feedback becomes an increasingly important mechanism for injecting momentum into the ISM, as the massive young stars produced are embedded in dense molecular environments to which they can effectively couple.

A star formation relation can be derived in the limit in which the ISM is supported against gravitational collapse by turbulent pressure \citep{Ostriker2011, Faucher-Giguere2013, Hayward2015, Torrey2016, Dib2017}.  Here, stellar feedback injects momentum into the ISM at a rate per area proportional to $\dot\Sigma_\star (P_\star / m_\star)$, where $(P_\star / m_\star)$ is the characteristic momentum injected per mass of young stars formed, and is dissipated in the mass of nearby gas per area $\Sigma_{\rm gas}$ on some characteristic timescale related to the coherence time of the turbulent eddies $t_{\rm eddy}$, where $t_{\rm eddy} \sim l_{\rm eddy}/\sigma_{\rm eddy}$, $l_{\rm eddy}$ being the spatial scale of the eddy and $\sigma_{\rm eddy}$ the turbulent velocity $\sigma_{\rm T}$.  As we are considering an approximately disk-like environment for star formation in the high gas surface density regime, the largest eddies will likely have length scales on the order of the disk scale height H \citep{Martizzi2016}, so $l_{\rm eddy} \sim \rm H \sim \sigma_{\rm T}/\Omega$, with $\sigma_{\rm T}$ being the turbulent velocity and $\Omega$ being the local orbital dynamical frequency.  We are concerned with the largest eddies, which contain most of the turbulent energy.  Hence, the timescales of turbulent energy dissipation scale as $t_{\rm eddy} \approx t_{\rm diss} \sim \rm H/\sigma_{\rm T} \sim \Omega^{-1}$.  Equating these rates of turbulent momentum injection and momentum dissipation in gas \footnote{{Here lies a direct connection to the no-feedback isolated disc simulations.  If star formation is equated to the mass flux of gas into a ``dense'' regime times a fixed efficiency, and that dense gas is then prevented from further star formation, we see that we expect a KS relation to arise with the correct slope and normalization, albeit a contrived one.}} , we find,
\begin{equation}
\dot\Sigma_\star (P_\star / m_\star) \approx \sigma_{\rm T} \Sigma_{\rm gas} / t_{\rm diss} \; , 
\end{equation}
substituting in our relations, this yields a star formation rate of,
\begin{equation} \label{rawturb}
 \dot\Sigma_\star \approx \sigma_{\rm T} \Omega \Sigma_{\rm gas} \left( \frac{P_\star}{m_\star} \right)^{-1} \; . 
\end{equation}

Relating $\sigma_{\rm T} \Omega$ back to the disk surface density with a modified Toomre-$Q$ parameter \citep{Toomre1964},
\begin{equation}\label{Q}
Q = \frac{\kappa \sqrt{c_s^2 + \sigma_{\rm T}^2 }}{\pi G \Sigma_{\rm disk}} \: ,
\end{equation}
where $\kappa$ is the epicyclic frequency $\sim \sqrt{2} \Omega$ for galactic potentials, $c_s$ is the sound speed, $\Sigma_{\rm disk} \approx \Sigma_{\rm gas} + \Sigma_\star$ is the disk surface density.  Here we include the self-gravity contribution from the collisionless stellar component of the disk, which is correct up to some order unity prefactor.  Assuming that we are turbulently rather than thermally supported, $ \sqrt{c_s^2 + \sigma_{\rm T}^2 } \approx  \sigma_{\rm T}$.  Substituting this in we find,
\begin{equation}
\dot\Sigma_\star \approx \frac{\pi}{\sqrt{2}} G Q   \left( \frac{P_\star}{m_\star} \right)^{-1} \Sigma_{\rm gas}(\Sigma_{\rm gas}+\Sigma_\star) \; . 
\end{equation}

Adopting a fiducial value for $(P_\star / m_\star)$ of $\sim 3000$ km s$^{-1}$ \citep[e.g.][]{Ostriker2011,Faucher-Giguere2013,Kim2015, Martizzi2015}, this yields
\begin{multline}\label{SF_eqn_high}
\dot\Sigma_\star \approx 3.3 \times 10^{-2} \left( \frac{P_\star / m_\star }{3000 \, {\rm km/s}} \right)^{-1} \left( \frac{\Sigma_{\rm gas}(\Sigma_{\rm gas}+\Sigma_\star)}{10^4 {\rm \, M_\odot^2 \, pc^{-4}}} \right) Q \\
 {\rm M_\odot \, yr^{-1} \, kpc^{-2}}\, . 
\end{multline}

For the gas-dominated regime, where $\Sigma_{\rm gas} \gg \Sigma_\star$, we recover a quadratic relation for star formation.  Similarly, should the stellar component dominate, as may be the case in stellar systems with older populations, a linear law in $\Sigma_{\rm gas}$ is found; this appears to be in good agreement with the slope of the KS relation seen in the FIRE runs.

The observed weak metallicity dependence seen in Figures \ref{rednmetals} \& \ref{metal_dep}, combined with the result shown in the lower left panel of Figure~\ref{fig:sf.z0}, which shows explicitly that the star formation rate varies with the strength of feedback, can be partly explained by a weak dependence of the final momentum injection from SNe feedback on the metallicity of the surrounding gas, e.g. $(P_\star / m_\star)^{-1} \sim Z^{0.114}$ \citep{Martizzi2015}.

The KS relation predicted by a turbulence-supported model assuming that the stellar surface density $\Sigma_\star \gg \Sigma_{\rm gas}$, with its approximately linear power-law slope, agrees remarkably well with the KS relation at moderately high surface density in the FIRE simulations.  Remarkably -- given the simplicity of the derivation -- when Equation \ref{SF_eqn_high} is used on a pixel-by-pixel basis to predict star formation rates from $\Sigma_{\rm HI + H_2}$ and $\Sigma_\star$, the predicted rates are nearly identical to the 10 Myr-averaged and instantaneous star formation rates, extending down to $\Sigma_{\rm HI + H_2} \approx 10^{-1}$ M$_\odot$ pc$^{-2}$.

\subsection{Low Gas Surface Density Regime}\label{disc:low}

%DIF < %%%%%%%%%% FIGURE 9 - Q vs. TAU ARGUMENT FIGURE
%DIF > %%%%%%%%%% FIGURE 10 - Q vs. TAU ARGUMENT FIGURE
\begin{figure}
	\centering
	\includegraphics[width=0.5\textwidth]{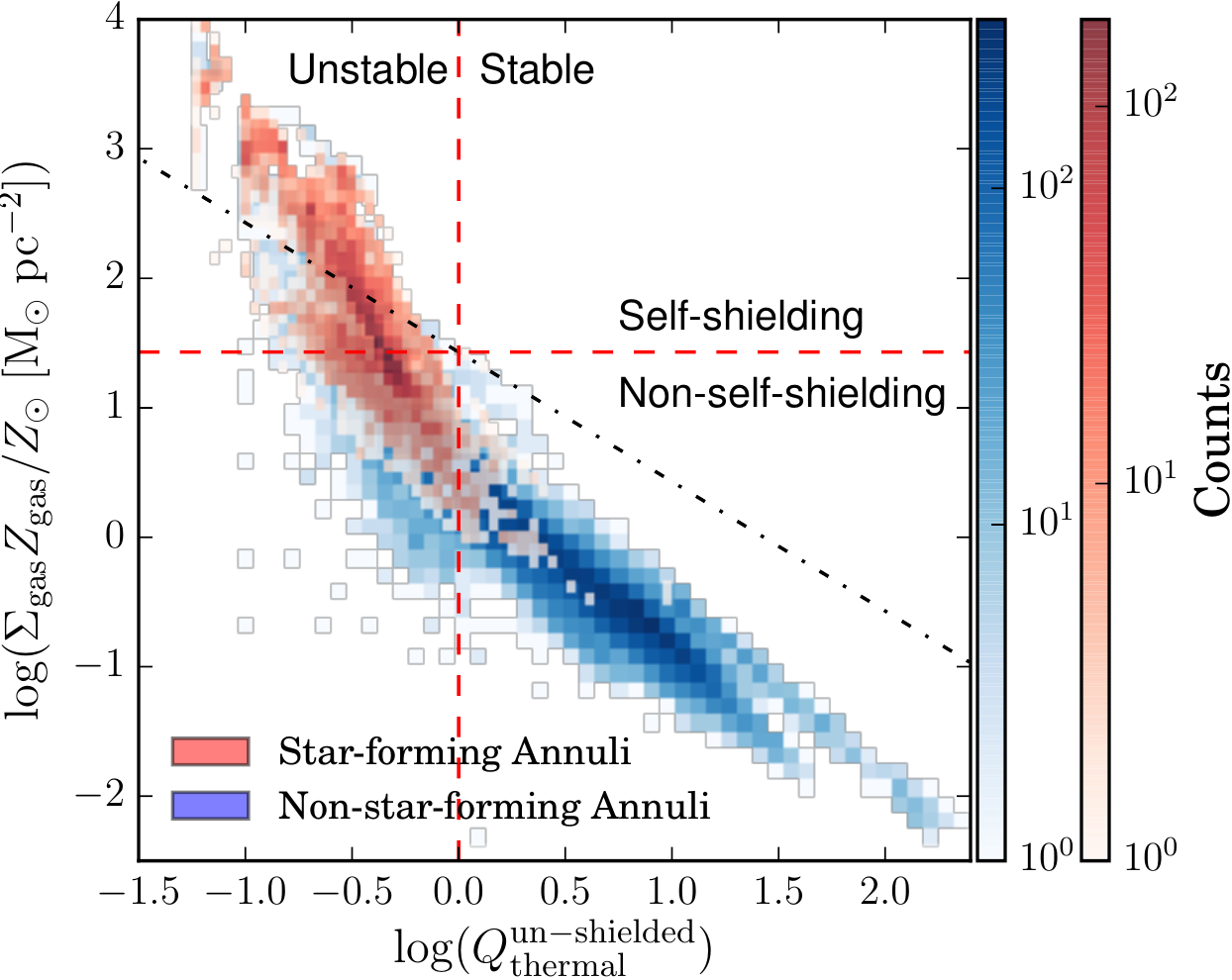}
	\caption{Comparison of whether self-shielding or gravitational instability determines the onset of efficient star formation (see Section \ref{thresholds}).  For each radial annulus in each galaxy (500 pc wide annuli, at each time from $z = 0.5 - 0$), we measure $\tilde{Q}_{\rm thermal}^{\rm un-shielded}$ (Eq. \ref{Qnoss}) and $\Sigma_{\rm gas} Z_{\rm gas}/Z_\odot$.  We plot a heat map of the number of pixels with each value of $\tilde{Q}_{\rm thermal}^{\rm un-shielded}$ and $\Sigma_{\rm gas}Z_{\rm gas}/Z_{\odot}$, color-coded so star-forming annuli are red (mean $\dot\Sigma_\star > 10^{-3}$ M$_\odot$ kpc$^{-2}$ yr$^{-1}$) and non-star-forming annuli are blue.  $\tilde{Q}_{\rm thermal}^{\rm un-shielded}$ is the Toomre-$Q$ parameter if the gas were purely thermally supported with $T =10^4$ K; this indicates whether the gas could be thermally stabilized against gravitational instabilities if it were not self-shielding. $\Sigma_{\rm gas}Z_{\rm gas}/Z_{\odot}$ is a proxy for optical depth, approximating whether or not the gas is self-shielding to ionizing radiation. Vertical and horizontal dotted red lines indicate the $Q=1$ stability threshold and the self-shielding threshold derived in \citet{Krumholz2009b}, respectively.  Black dotted line shows the track of varying $\Sigma_{\rm gas}$ at fixed $Z'$ and $\Omega$.  The onset of star formation clearly occurs around $\tilde{Q}_{\rm thermal}^{\rm un-shielded} \sim 1$, even though the annuli are not self-shielding-- i.e. gravitational instability initiates collapse, which then produces dense self-shielding clumps.}
	\label{toomreQ}
\end{figure}
%DIF < %%%%%%%%%% FIGURE 10 - PHYSICS TESTS PANEL
%DIF > %%%%%%%%%% FIGURE 11 - PHYSICS TESTS PANEL
\begin{figure}
	\centering
	\includegraphics[width=0.44\textwidth]{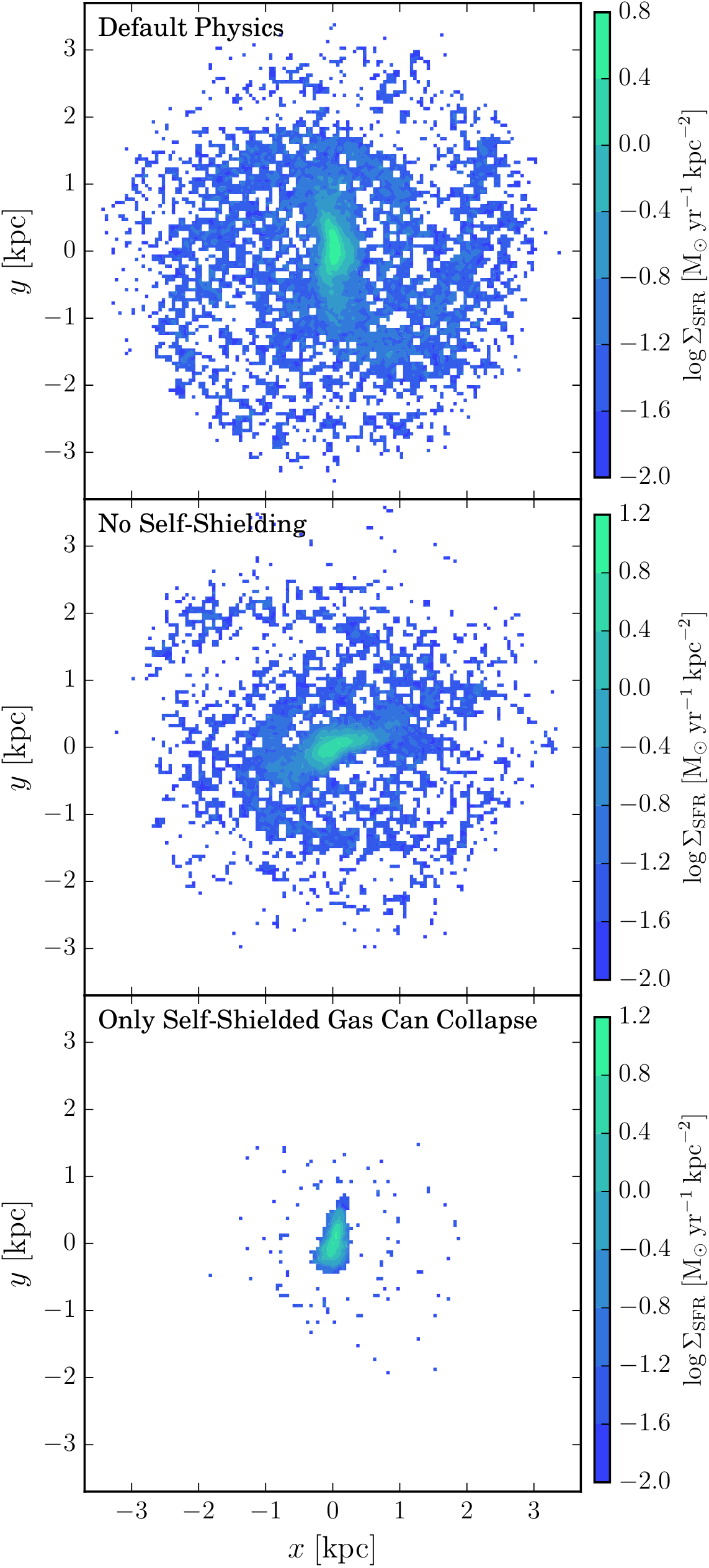}
	\caption{Demonstration of the importance of gravitational instability versus self-shielding in star formation.  One Milky Way-mass simulation, {\bf m12i} presented in \citet{Hopkins2014}, was restarted from $ z \approx 0.07$ and rerun with three sets of physics, the default physics implemented in the FIRE runs ({\bf top}), one with both self-shielding and cooling below $10^4$ K disabled ({\bf middle}), and one where shielded gas had normal properties but non-shielded gas had a large artificial pressure floor, effectively disabling gravitational fragmentation until the gas first became self-shielding ({\bf bottom}).  Colored pixels indicate the extent and intensity of star formation since the runs were restarted in the form of the 500 Myr averaged star formation rate surface density. The star formation largely has the same structure with or without self-shielding, as long as fragmentation is allowed (gas can still isothermally collapse and meet the star formation criteria).  But when only annuli in a galaxy which are entirely self-shielding can fragment or collapse, only the central $\sim$kpc of the galaxy (where the gas is entirely molecular) efficiently forms stars.}
	\label{phystests}
\end{figure}

At the other extreme of galactic environments, we consider the low gas surface density regime in which gas is supported by thermal -- rather than turbulent -- pressure. We expect this transition to occur for $\Sigma_{\rm gas} \lesssim 10$~M$_\odot$~pc$^{-2}$ \citep[see][for details]{Schaye2004, Ostriker2010, Hayward2015}.  In this regime, a star formation equilibrium rate can be derived by balancing photoheating from young stars with gas cooling.  At extremely low gas surface densities, where $\Sigma_{\rm gas} \ll 1 $ M$_\odot$ pc$^{-2}$, the metagalactic UV background itself may become the predominant source of heating, requiring no star formation at all to maintain a thermal pressure equilibrium, providing a physical SFR floor \citep{Schaye2004, Ostriker2010}.

As derived in \citet{Ostriker2010} as an ``outer-disk'' law, we can balance photoheating with radiative gas cooling.  For ionizing and photo-electric photons dominating the gas heating, the heating rate per area is
\begin{equation}
\frac{\rm \dot E_{\rm heat}}{l^2} = \frac{f_{\rm abs}\beta \rm L_\star}{l^2} = f_{\rm abs}\beta \epsilon c^2 \dot\Sigma_\star \: ,
\end{equation}
where $f_{\rm abs} (\lesssim 1)$ is the fraction of the emitted photoheating photons absorbed by surrounding gas, $\beta \sim 0.1$ is the fraction of ionizing radiation emitted by young stars \citep{Leitherer1999}, and $\epsilon \sim 4\times 10^{-4}$ is the fraction of rest-mass energy radiated by stars in their lifetimes.  On the other hand, the cooling rate per area is
\begin{equation}
\frac{\rm \dot E_{\rm cool}}{l^2} = \frac{\Lambda n_e n_i V \rm}{l^2} \approx \frac{\Lambda Z n_g \Sigma_{\rm gas} \rm}{\mu} \: ,
\end{equation}
with $\Lambda \sim 10^{-22}$ erg s$^{-1}$ cm$^{-3}$ being the net cooling rate \citep{Robertson2008}; $n_e$, $n_i$, and $n_g$ being the electron, ion, and gas number densities; and $V \sim l^2 h$ being the volume of gas considered.  Equating the heating and cooling rates, we find
 \begin{equation}
\dot\Sigma_\star \approx \frac{\Lambda Z n_g \Sigma_{\rm gas} \rm}{f_{\rm abs} \mu \beta \epsilon c^2} \: .
\end{equation}
Furthermore, we have $n_g = \rho_{\rm gas}/\mu \approx \Sigma_{\rm gas}/2h\mu$ and $h~\approx~c_s/\Omega$ in the thermally supported limit, as $\sqrt{c_s^2 + \sigma_T^2} \approx c_s$.  Thus, we have $n_g \approx \Sigma_{\rm gas} \Omega/2 c_s \mu$, and $\dot\Sigma_\star$ becomes
 \begin{equation} \label{rawtherm}
 \dot\Sigma_\star \approx \frac{\Lambda Z \Omega \Sigma_{\rm gas}^2 \rm}{2 f_{\rm abs} \mu^2 \beta \epsilon c^2 c_s} = \frac{\Sigma_0 Z \Omega}{f_{\rm abs}} \left( \frac{\Sigma_{\rm gas}}{\Sigma_0} \right)^2 \: ,
\end{equation}
where $\Sigma_0 \approx 2 \mu^2 \beta \epsilon c^2 c_s / \Lambda \approx 4$ M$_\odot$ pc$^{-2}$, assuming $T = 10^4$~K, for which $c_s \approx 12$ km s$^{-1}$.  Scaling this to approximately Milky-Way values ($\Omega \approx v_c/R \approx 220$  km s$^{-1}$/R, $Z \approx Z_\odot$), we have
\begin{multline}\label{SF_eqn_low}
\dot\Sigma_\star \approx 1.3\times 10^{-3} \left( \frac{Z}{Z_\odot}\right)  \left( \frac{10 \; \rm kpc}{R} \right) \left( \frac{\Sigma_{\rm gas}}{\Sigma_0} \right)^2 \frac{1}{f_{\rm abs}}  \\
 {\rm M_\odot \, yr^{-1} \, kpc^{-2}}\, .
\end{multline}

The star formation rate relation for a thermally supported ISM has the same $\dot\Sigma_\star \propto \Sigma_{\rm gas}^2$ dependence as the turbulently supported high gas surface density regime when the gas surface density dominates, but with an added dependence on $Z$ and $\Omega$.  This is similar to the relations found by \citet{Ostriker2010} and \citet{Krumholz2009b} at low surface densities, with their dependence on metallicity.  The scaling here is in good agreement with the FIRE runs at low gas surface densities for the $\Sigma_{\rm HI + H_2}$ tracer but differs from the shallower relations found by some observational studies \citep{Bigiel2010, Roychowdhury2015}.  For low column densities, where $f_{\rm abs} \ll 1$ (and usually $f_g \ll 1$), the fraction of absorbed photoheating photons may go as the optical depth and thus the gas surface density $f_{\rm abs} \propto (1-\exp{(-\tau)}) \approx \tau \propto \Sigma_{\rm gas} Z'$, reducing the low gas surface density relation to $\dot\Sigma_\star \propto \Sigma_{\rm gas}$, degenerate in form with the relation derived for the high gas surface density regime when $f_g \ll 1$, which may explain the aforementioned shallower observations and weak $Z'$ dependence seen in Figure~\ref{metal_dep}.  Similarly to the turbulently supported regime derivation, comparing the predictive ability of Equation \ref{SF_eqn_low} using pixel-by-pixel values of $Z$, $\Omega$, and $\Sigma_{\rm HI + H_2}$ with the measured star formation rates, very close agreement is found for $\Sigma_{\rm HI + H_2} \lesssim 10$ M$_\odot$ pc$^{-2}$, the regime in which the relation is expected to apply \citep{Hayward2015}.

The transition between linear and quadratic dependences of the star formation rates on the gas surface density in the KS relation, for both the low and high gas surface density regimes may explain some of the `kinks' seen in the relation by various observers \citep[e.g.][]{Bigiel2008}.

\subsection{Star Formation Thresholds}\label{thresholds}

There remains the question of what, physically, {\em fires} up star formation in the simulations in the first place.  Figures~\ref{toomreQ} and \ref{phystests} address this question. 

Consider a radial annulus of a smooth gas disk at some large radius $R$.  At sufficiently large $R$ and low densities, the disk is not self-shielding to UV radiation, and even the meta-galactic UV background is sufficient to maintain the disk at warm temperatures $T\sim 10^{4}$\,K. The thermal Toomre-$Q$ parameter at this temperature: 
\begin{align}\label{Qnoss}
\tilde{Q}_{\rm thermal}^{\rm un-shielded} \equiv \frac{\kappa\,c_{s}(10^{4}\, \rm K)}{\pi\,G\,\Sigma_{\rm disk}} \approx 1.2\,\left( \frac{\Omega}{{\rm Gyr^{-1}}} \right)\,\left( \frac{\rm M_{\sun}\,{\rm pc^{-2}}}{\Sigma_{\rm disk}} \right)
\end{align}
is $\tilde{Q}_{\rm thermal}^{\rm un-shielded} \gg 1$, i.e.\ the disk is fully stable. In this limit, we do not expect (nor see in our simulations) any significant star formation. In the opposite limit (in e.g.\ the centers of massive galaxies), the surface densities are high, $\tilde{Q}_{\rm thermal}^{\rm un-shielded} \ll 1$ (i.e.\ thermal support, even in the warm gas, is insufficient to stabilize the disk, so it is supersonically turbulent, with $Q\sim Q_{\rm turb} \sim 1$) and the gas is self-shielding. In this limit, of course, we see efficient star formation.

What determines the transition between these two limits? It has been suggested that when an annulus reaches a critical column density to become self-shielding, it can suddenly cool to $T\ll 10^{4}\,$K, reducing the thermal pressure support of the gas disk against fragmentation (i.e.\ lowering the Toomre-$Q$ from $\gg1$ to $\ll 1$) and so initiating gravitational collapse and star formation \citep[e.g.][]{Schaye2004, Krumholz2009a}. This will occur when the dust optical depth $\tau = \Sigma_{\rm gas}\,\kappa$ exceeds unity, or more accurately from \citet{Krumholz2009a}, when $\Sigma_{\rm gas}\,Z^{\prime} > 27\,\rm M_\odot\,{\rm pc^{-2}}$ (where $Z^{\prime}\equiv Z/Z_{\sun}$ reflects the assumption of a constant dust-to-metals ratio). 

Alternatively, an annulus which is {\em not} self-shielding (hence at a temperature $T\sim 10^{4}\,$K) will still become gravitationally unstable, when $\tilde{Q}_{\rm thermal}^{\rm un-shielded} \lesssim 1$, i.e.\ $\Sigma_{\rm disk}\,(\Omega/{\rm Gyr}^{-1})^{-1} \gtrsim 0.7\,\rm M_{\sun}\,{\rm pc}^{-2}$. The annulus would then rapidly fragment isothermally (at $\sim10^{4}\,$K) at first, until individual overdensities/fragments quickly become internally self-shielding (reaching {\em local} surface densities $\Sigma_{\rm gas}\,Z^{\prime} > 27\,\rm M_\odot\,{\rm pc^{-2}}$ as they collapse), then cool and fragment further to form stars. During this collapse, super-sonic turbulence would be driven by gravitational instabilities and feedback to maintain a turbulent $Q\sim1$, but the important point is that the {\em thermal} support ($\tilde{Q}_{\rm thermal}^{\rm un-shielded}$) is insufficient.

The question is essentially which of these thresholds is reached ``first.'' Figure~8 examines this in our simulations by plotting all annuli in the space of $\tilde{Q}_{\rm thermal}^{\rm un-shielded} \propto \Omega/\Sigma_{\rm disk}$ versus $\tau_{\rm shielding} \propto \Sigma_{\rm gas}\,Z$, and identifying those which are and are not star-forming. Clearly, robust star formation occurs in annuli which are not, on average, self-shielding (they have $\Sigma_{\rm gas}\,Z^{\prime} \approx 1-5\,\rm M_\odot\,{\rm pc^{-2}}$). We stress that the small sub-regions where star-formation is occurring {\em within} those annuli are of course self-shielding (this is in fact required by our resolution-scale star formation model), and reach $\Sigma_{\rm gas}\,Z^{\prime}\gg 100\,\rm M_\odot\,{\rm pc^{-2}}$ locally.  In contrast, the onset of star formation corresponds very closely to where $\tilde{Q}_{\rm thermal}^{\rm un-shielded}\approx 1$. This is consistent with observations of star-forming spiral galaxies by \citet{Martin2001} who found that gravitational instability thresholds were sufficient to explain the extent of star-forming disks.

Examining Figure \ref{toomreQ} further, we see that the annuli all lie on a track which intercepts the instability threshold $\tilde{Q}_{\rm thermal}^{\rm un-shielded} \approx 1$ more than a dex below the self-shielding threshold $\Sigma_{\rm gas}Z' \sim 27$ M$_\odot$ pc$^{-2}$.  Star formation (red pixels) is seen as annuli cross the instability line, and the distribution then turns upwards as star formation begins to enrich the annuli in $Z'$ without much affecting $\Sigma_{\rm gas}$ or $\tilde{Q}_{\rm thermal}^{\rm un-shielded}$ instantaneously.  As annuli cross into the self-shielding regime, $\Sigma_{\rm gas}$ quickly crosses into the high-surface-density regime, and vigorous star formation results in short depletion timescales for these annuli, preventing highly shielded, low-$\tilde{Q}_{\rm thermal}^{\rm un-shielded}$ annuli from remaining long in that regime.  Crossing into the self-shielded regime appears to coincide with the rapid rise in the lower envelope of $\dot\Sigma_\star$-- these regions are vigorously forming stars throughout and are unable to prevent themselves from cooling rapidly and fragmenting, as in \citet{Schaye2004}.  Moreover, the high-$\tilde{Q}_{\rm thermal}^{\rm un-shielded}$ $\gg 10$ annuli with very low $\Sigma_{\rm gas} Z'$ appear to come from the galactic outskirts at several times the half-mass radii.

To verify the relative importance of gravitational instability versus self-shielding, we also considered an idealized numerical experiment in Figure~\ref{phystests}. Specifically, we took one of our Milky Way-mass galaxy simulations (run {\bf m12i} from \citealt{Hopkins2014}) and re-ran it for about $\sim1$\,Gyr close to $z=0$ (from $z=0.07$ to $z=0$), modifying the physics in the re-run. We considered two cases. 
\begin{itemize}
\item{\bf (1)} ``No Self-Shielding:'' In this case we disable self-shielding in our radiative heating routines and do not allow any cooling below $10^{4}\,$K.\footnote{In our ``no self-shielding'' run, we still enforce the \citet{Krumholz2009b} requirement described in \S~2 for whether an individual gas particle is allowed to form stars, since this is (strictly speaking) just a metallicity-dependent {\em local} surface density threshold ($\Sigma_{\rm gas}\,Z^{\prime} > 27\,\rm M_\odot\,{\rm pc^{-2}}$) evaluated at the resolution scale.} Clearly, Figure~\ref{phystests} shows that gas is still able to fragment and form stars -- the spatial extent of the star formation is nearly identical to our ``default'' run, in fact, indicating that cooling to $T\ll 10^{4}\,$K is not what determines the outer cutoff of star formation in the disk (consistent with our argument in Figure~\ref{toomreQ}). The total star formation rate is also similar within $15\%$.

\item{\bf (2)} ``Only Self-Shielded Gas Can Collapse:'' If self-shielding always preceded fragmentation and star formation, we should be able to disable Toomre-style fragmentation in gas which is not self-shielding, and obtain the same result. This is non-trivial in practice. We attempt to implement this as follows: for gas which is self-shielding (has cooled to $<8000\,$K and/or meets the \citet{Krumholz2009b} criterion), the physics is ``normal,'' but for gas which is not self-shielding, we add an artificial pressure term to the hydrodynamic equations ($P \rightarrow P_{\rm true} + P_{\rm floor}$) where $P_{\rm floor} = 4\times10^{-11}\,(n/{\rm cm^{-3}})$ (i.e. the pressure the gas would have at $3\times 10^{5}$\,K). The specific value is chosen to ensure the non-shielded gas has an ``effective'' Toomre-$Q \sim $ a few (sufficient to prevent fragmentation but not ``blow up'' the galaxy). When we do this, we see that efficient star formation becomes restricted to the central $\sim$\,kpc only (and the total star formation rate falls by a factor $\approx 3$). This central region is basically the location of the molecular disk -- i.e.\ the regime where the gas is {\em entirely} molecular, since that is where the disk is uniformly self-shielding. Clearly, this is not a good description of star formation in the ``default'' simulation.
\end{itemize}

\subsubsection{Star Formation in the Small Magellanic Cloud}\label{magellanic}
The star formation threshold behavior seen in our simulations is consistent with observations of the SMC, as measured by \citet{Bolatto2011} and \citet{Hony2015}.  The star formation rates seen by \citet{Hony2015} agree well with the spatially resolved KS relation when considering young star counts as a measure of $\dot\Sigma_\star$.  Moreover, considering the metallicity and surface density of the SMC, in the SMC body/wing, $\tau \sim \Sigma_{\rm gas} Z/Z_\odot \sim 10$ M$_\odot$ pc$^{-2}$, and in the SMC tail, $ \Sigma_{\rm gas} Z/Z_\odot \sim 2$ M$_\odot$ pc$^{-2}$ \citep{Oliveira2009, Nidever2008}, the SMC body is thus not quite at the self-shielding threshold, and  the tail is certainly not.  On the other hand, when estimating $Q$ for the SMC wing, one finds $Q \sim 2/3$ and for the body $Q \ll 1$ (assuming a linearly rising rotation curve to $v_{\rm rot} \sim 50$ km s$^{-1}$ at $R \sim 3.5$ kpc found by \citet{Stanimirovic2004}); hence, the SMC appears to be consistent with gravitational instabilities triggering star formation.
%------------------------------------------------------------------------------------
\section{Conclusions}\label{cons}

In this paper, we investigated the spatially-resolved KS relation in the cosmological FIRE simulations with $z = 0$ central halo masses ranging from $10^{10}$ M$_\odot$ to $10^{13}$ M$_\odot$. Our principal conclusions are the following:
\begin{itemize}
\item \rev{The simulated galaxies exhibit a KS-like relation with slope and scatter consistent with observations. We emphasize that this relation emerges naturally rather than being imposed `by hand' because we assume an instantaneous local star formation efficiency of 100 per cent per free-fall time, but the predicted global star formation efficiency is low, consistent with observations, due to stellar feedback. This is true across many orders of magnitude in halo mass and at all redshifts studied. The exact slope of the relationship between $\Sigma_{\rm SFR}$ and $\Sigma_{\rm gas}$ depends on the gas and star formation tracers used, but the two quantities are tightly correlated in all cases explored.}

\item \rev{For the neutral hydrogen form of the relation, the simulations are observations agree well also in terms of normalisation. In contrast, for the molecular hydrogen relation, the simulated and observed relations are systematically offset, likely because our proxy for the molecular gas (the `Cold \& Dense' tracer) underestimates the true molecular gas mass by $\sim 0.5$ dex for gas surface density $\lesssim 100 M_{\odot}$ pc$^{-2}$.}

\item The time-averaged KS relation does not appear to have a significant dependence on pixel size (i.e. map resolution) for gas surface densities with sufficiently resolved star formation rate distributions (i.e. above the $\Sigma_{\rm gas}$ where the KS relation would yield at least a few young star particles per pixel given our mass resolution; see Section \ref{meth}), with the slope of the power law remaining effectively unchanged.  However, we are unable to resolve star formation rates at gas surface densities at our smallest pixel size (100 pc), for which observations exhibited large scatter in the KS relation ($\Sigma_{\rm gas} \lesssim 10$ M$_\odot$ pc$^{-2}$).

\item The KS relation and star formation efficiency in the FIRE simulations is independent of redshift.  The simulations do not exhibit any metallicity-dependent cutoff; however, the star formation rate surface density is weakly dependent on the metallicity, on the order expected from SNe feedback's momentum injection dependence on metallicity \citep{Cioffi1988, Martizzi2015}.

\item{At the high end of gas surface density, where $\Sigma_{\rm gas} \gtrsim 100$ M$_\odot$ pc$^2$ and gas is predominantly molecular, we find that the KS relation obeyed by the simulated galaxies is consistent with injection of momentum from supernovae balancing momentum dissipation in turbulence, or analogously, turbulent ``pressure'' maintaining vertical hydrostatic equilibrium \citep{Ostriker2011,Faucher-Giguere2013,Hayward2015}.  This explanation yields a power law independent of redshift or metallicity at high gas surface densities, where $\dot\Sigma_\star \propto \Sigma_{\rm gas}\Sigma_{\rm disk}$ (see \S4.1).  Because the disks in our simulations are not particularly gas rich, we find a slightly steeper than linear KS relation in this regime.}

\item In regions of low gas surface density ($\Sigma_{\rm gas} \lesssim 10$ M$_\odot$ pc$^{-2}$), characteristically in galaxy outskirts and regions between spiral arms, our spatially-resolved KS relation agrees well with that expected from a simple local equilibrium between photoheating from ionizing, or near-ionizing, radiation from young stars and radiative gas cooling.  This argument yields a $\dot\Sigma_\star \propto Z \Sigma_{\rm gas}^2$ power law (see \S4.2), as discussed in \citet{Ostriker2010} and \citet{Hayward2015}.

\item  Vigorous star formation begins as gas self-gravity overcomes the gas thermal pressure gradient, thus making the gas Toomre-unstable.  This self-gravity driven collapse occurs around $\Sigma_{\rm gas} \sim 1$ M$_\odot$ pc$^{-2}$, an order of magnitude before the gas becomes self-shielding to UV radiation, at $\Sigma_{\rm gas} \sim 27$ M$_\odot$ pc$^{-2}$.  Thus, we find that in the FIRE simulations, star formation is triggered by gravitational instabilities, which then cause the gas to fragment and collapse, thereby becoming self-shielding to ionizing radiation, cooling rapidly and forming stars.  The threshold for gravitational instability, $Q \sim \Omega/ \Sigma_{\rm disk}$, depends only on the density of gas and stars, i.e. the criterion for warm gas ($T \gtrsim 10^4$ K) to support itself thermally against fragmentation, and subsequently star formation, is independent of both $Z$ and $z$.
\end{itemize}

Future observations with high spatial resolution and sensitivity to low surface brightnesses should aid in understanding the outskirts of galactic environments where star formation is on the brink of \emph{firing} up and the surface densities of gas and stars are near the thresholds of gravitational instability and self-shielding.  This will help determine if gravitational fragmentation -- rather than self-shielding -- is indeed the primary triggering mechanism of star formation. {Similarly, future work implementing chemical networks and radiative transfer post-processing in cosmological simulations will help bridge the gap between simulated tracers of star formation and molecular gas and resolved observations.} 

%------------------------------------------------------------------------------------
\section*{Acknowledgments}
MEO is grateful for the encouragement of his late father, SRO, in studying astrophysics, and for many helpful discussions with A. Wetzel, J. Schaye, S. Dib, and I. Escala. We are grateful to the anonymous referee for providing us with constructive comments and suggestions, which have significantly improved the work. This research has made use of NASA's Astrophysics Data System.  MEO was supported by the National Science Foundation Graduate Research Fellowship under Grant No. 1144469.  CCH is grateful to the Gordon and Betty Moore Foundation for financial support.  The Flatiron Institute is supported by the Simons Foundation.  Support for PFH was provided by an Alfred P. Sloan Research Fellowship, NASA ATP Grant NNX14AH35G, and NSF Collaborative Research Grant \#1411920 and CAREER grant \#1455342. Numerical calculations were run on the Caltech compute cluster ``Zwicky'' (NSF MRI award \#PHY-0960291) and allocations TG-AST120025, and TG-AST130039 granted by the Extreme Science and Engineering Discovery Environment (XSEDE) supported by the NSF. CAFG was supported by NSF through grants AST-1412836 and AST-1517491, by NASA through grant NNX15AB22G, and by STScI through grants HST-AR-14293.001-A and HST-GO-14268.022-A.  {RF was supported by the Swiss National Science Foundation (Grant No. 157591).} DK acknowledges support from the NSF grant AST-1412153 and Cottrell Scholar Award from the Research Corporation for Science Advancement. EQ was supported by NASA ATP grant 12-ATP12-0183, a Simons Investigator award from the Simons Foundation, and the David and Lucile Packard Foundation.

% Set up the bib the way I did the paper with P&J
\bibliography{KS_paper}
\bibliographystyle{mnras}

%==========================APPENDIX==============================
\appendix
\section{Robustness of Star Formation Rates to Variations in Star Formation, Cooling, and Stellar Feedback}
\label{sec:appendix:baryonic.physics}

%%%%%%%%%%% FIGURE A1 - SF LAWS TEST PANEL
\begin{figure*}
	\centering
	\includegraphics[width=0.83\textwidth]{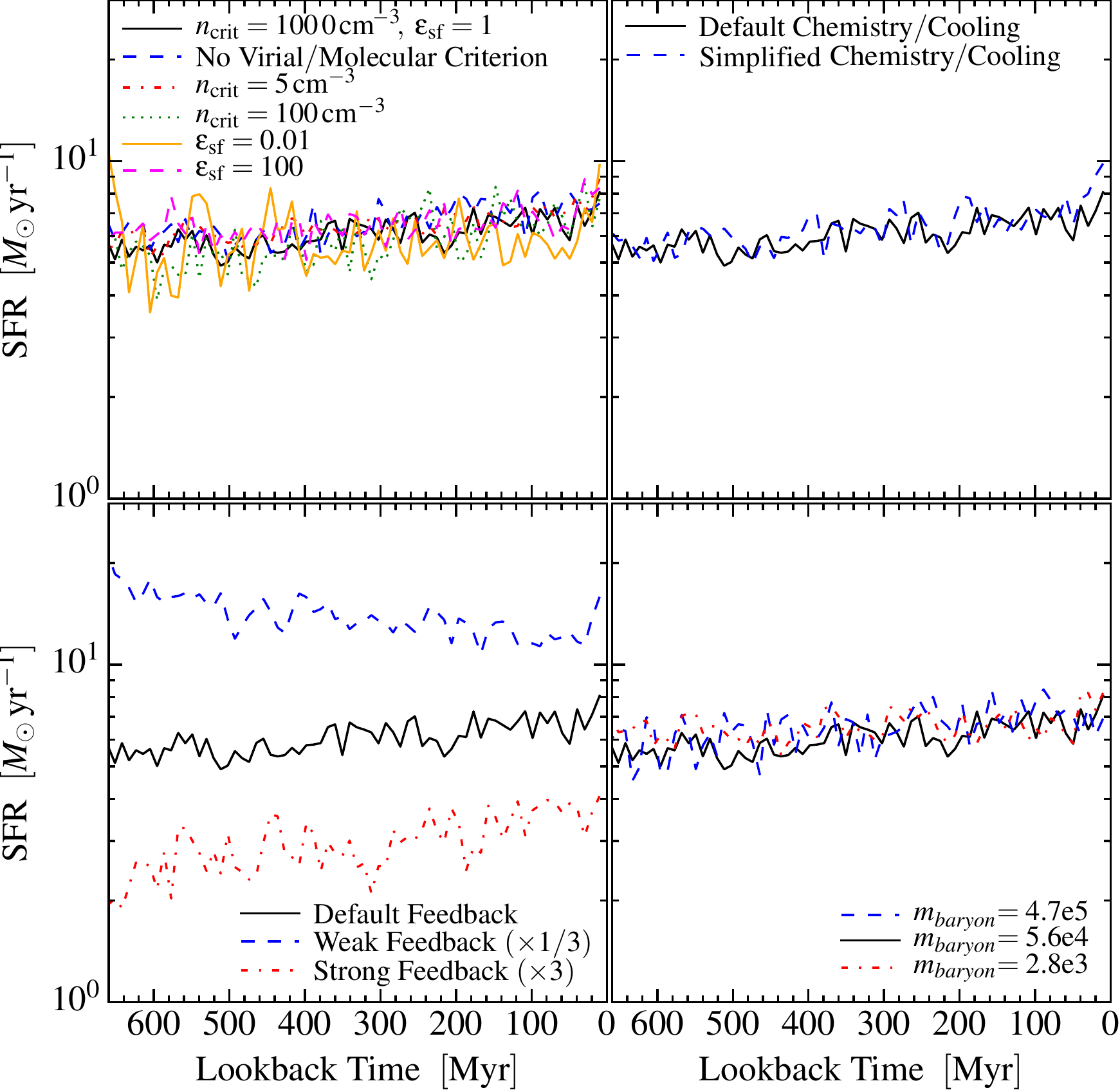}
   	\vspace{-0.25cm}
    	\caption{Star formation rate versus time in our MW-mass ({\bf m12i}) simulation from redshift $z\approx 0.07-0$; the simulation was restarted at $z = 0.07$ and run with varying parameters to study the effect on the star formation rate given the same initial galaxy properties. 
    {\bf Top Left:} Effect of the resolution-scale star formation criteria. In our ``default'' model, gas that is self-gravitating, molecular, and dense ($n>n_{\rm crit}=1000\,{\rm cm^{-3}}$) forms stars at a rate $\dot{\rho} = \epsilon_{\rm sf}\,\rho_{\rm mol}/t_{\rm ff}$, with $\epsilon_{\rm sf}=1$. We compare (1) removing the self-gravity \&\ molecular restrictions, (2-3) varying $n_{\rm crit}$, and (4-5) varying $\epsilon_{\rm sf}$.
    {\bf Bottom Left:} We vary the strength of feedback by multiplying/dividing the rates of all mechanisms per unit stellar mass by $3$ relative to the predictions from the stellar evolution models.
    {\bf Top Right:} Default physical cooling model compared with a toy model that ignores all low-temperature cooling physics and puts all gas on a single, solar-metallicity cooling curve.
    {\bf Bottom Right:} Resolution effects, changing the baryonic particle mass. 
    The results of this figure show that only the strength of feedback significantly alters the star formation rate at fixed $\Sigma_{\rm gas}$, varying the sub-grid star formation law has essentially no effect.
    \label{fig:sf.z0}}
\end{figure*}
%%%%%%%%%%% FIGURE A2 - SF LAWS NUMBER DENSITY CDFs
\begin{figure}
	\centering
	\includegraphics[width=0.47\textwidth]{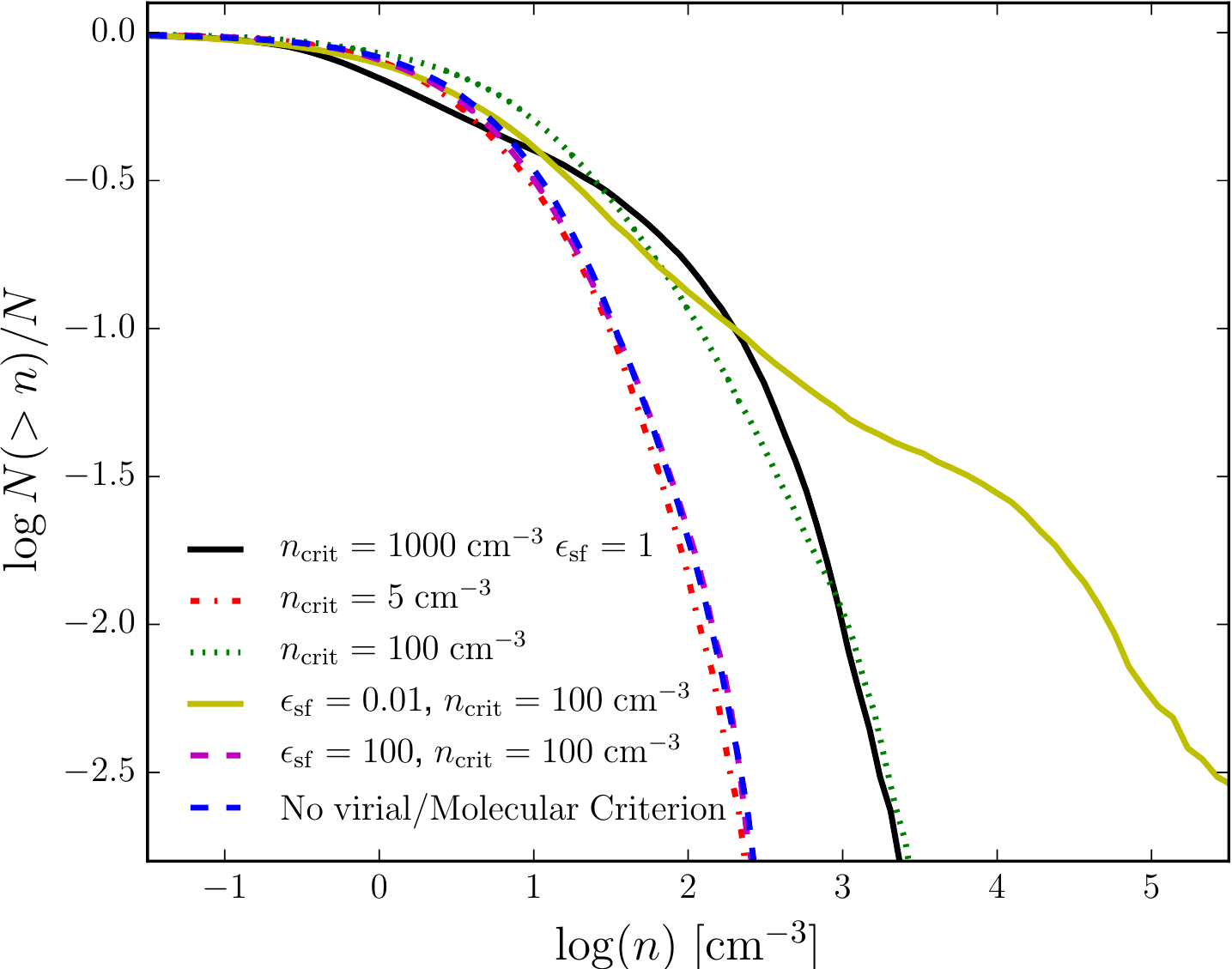}
   	\vspace{-0.25cm}
    	\caption{Gas number density CDF in our MW-mass ({\bf m12i}) simulation at redshift $z\approx 0$ for the various star formation model test runs in the upper left panel of Figure~\ref{fig:sf.z0}.  The CDFs evolve dynamically such that the ``correct'' amount of dense gas forms to support the required SFR to regulate the galaxy.  In our (new) ``default'' model, with the highest $n_{\rm crit}=1000\,{\rm cm^{-3}}$ more gas evolves to higher densities before turning into stars, compared to the other SF models except for our low-efficiency run.  Removing the virial and molecular thresholds is nearly equivalent to drastically reducing $n_{\rm crit}$ (it had $n_{\rm crit}=100$ cm$^{-3}$) or increasing $\epsilon$.  Intuitively, holding $\dot\rho_\star$ constant in Eq.~\ref{eq:sfr}, i.e. the SFRs converge to the `necessary' value, we expect that $n \propto \epsilon^{-1/2}$.  Indeed, we see that increasing $\epsilon$ by a factor of 100 moves the gas density CDF a dex towards lower densities.}
    \label{densCDFs}
\end{figure}

Here, we demonstrate the robustness of the star formation rate in the FIRE simulations to reasonable changes in the implemented star formation, cooling, and stellar feedback physics \citep[reviewed in detail in][]{Hopkins2014}.  A number of previous studies have consistently demonstrated the convergence of star formation rates and the KS relation, with resolution and numerical implementations of star formation \citep[][]{Saitoh2008, Federrath2012, Hopkins2012b, Hopkins2013d, Hopkins2013b, Hopkins2013c, Hopkins2016, Agertz2013}.  In Figure~\ref{fig:sf.z0}, we illustrate this with a set of simple tests using the newest version of the code \citep[part of a more general numerical study, presented in detail in][]{Hopkins2017}.  In each case, we re-start the same Milky-Way mass simulation \citep[{\bf m12i} from][as in Section \ref{thresholds} in the text]{Hopkins2014}, and re-run it from $z = 0.07 - 0$ with different numerical choices.  This ensures the initial conditions are identical; $\Sigma_{\rm gas}$ for the galaxy, for example, is fixed, so we can simply read off from the star formation rate whether the galaxy's location in the KS law would change.

We compare our default star formation model, using the criteria enumerated in Section \ref{meth}, here with $n_{\rm crit}=1000\,{\rm cm^{-3}}$ and $\epsilon_{\rm sf}=1$, where $\epsilon_{\rm sf}$ represents the local efficiency with which gas turns into stars in a free fall time, i.e. $\dot\rho_\star = \epsilon_{\rm sf} \rho_{\rm mol}/t_{\rm ff}$, to variations with $\epsilon_{\rm sf}=0.01-100$, $n_{\rm crit}=5-1000\,{\rm cm^{-3}}$, and turning on/off the additional virial and molecular criteria. We find that the star formation rate (and indeed the entire spatially-resolved KS relation) is effectively the same in all cases.  

The gas in the restarts responds dynamically to these changes in the star formation prescription, as seen in Figure~\ref{densCDFs}.  Variations in $n_{\rm crit}$ allow gas to evolve to higher/lower densities before turning into stars rapidly, seen in the rapid fall-off of the densest gas in the various models.  Changes in $\epsilon_{\rm sf}$ yield similar results, with smaller $\epsilon_{\rm sf}$ values allowing gas to continue evolving to higher physical densities.  Removing the virial and molecular criteria appears to have the same effect as lowering the density threshold or raising the local star formation efficiency, likely as more of the gas just above the threshold is converted to stars rapidly that is not necessarily bound. All the while, the star formation rates in the restarts are essentially unchanged; we see that the gas in the galaxies is dynamically evolving to produce the `correct' star formation rate to regulate itself.  Detailed observations of the gas density CDF in the Milky Way and nearby galaxies may thus help constrain sub-grid star formation prescriptions to produce realistic gas density distributions, without altering the overall star formation rates in the simulations.

In Figure~\ref{fig:sf.z0}, we also vary our cooling model, replacing all low-temperature cooling physics with a single cooling rate, putting all gas on a single, solar-metallicity cooling curve, and removing the molecular star formation requirement. We see that there is no effect on the star formation rate; we similarly find no effect on the outflow rate or global morphology. Details of the phase structure, of course, differ, but these have no {\em large dynamical effect}, consistent with various previous studies that have found that almost all gas in galaxies is super-sonically turbulent and has cooling times much shorter than their dynamical times \citep{Hopkins2011, Hopkins2012a, Glover2012}.

We also explicitly consider the mass resolution convergence by up- and down-sampling the particle distribution with particle splitting/merging. We find that the star formation rate is nearly identical over $\sim 2.5$\,dex in mass resolution, even a factor of $\sim 10$ lower resolution compared to our ``standard FIRE'' resolution. This is consistent with our argument in Section \ref{meth} that we only need to marginally resolve the Toomre scale to achieve convergence in the star formation rate because the most massive clouds dominate star formation \citep{Williams1997}.

As seen in the bottom left panel of Figure~\ref{fig:sf.z0}, variation in the strength of feedback per mass of young stars is the only effective means of changing the star formation rate.  As galaxies self-regulate for a given level of feedback, changing the strength of feedback systematically results in higher star formation rates for lower levels of feedback per star and vice versa.  This is consistent with our scalings in Section \ref{disc}.

\section{Various Molecular Gas Mass Proxies in the FIRE Simulations}
\label{sec:appendix:molefrac}

%%%% FIGURE B1: Various Molecular Proxies for KS.
\begin{figure}
	\centering
	\includegraphics[width=0.47\textwidth]{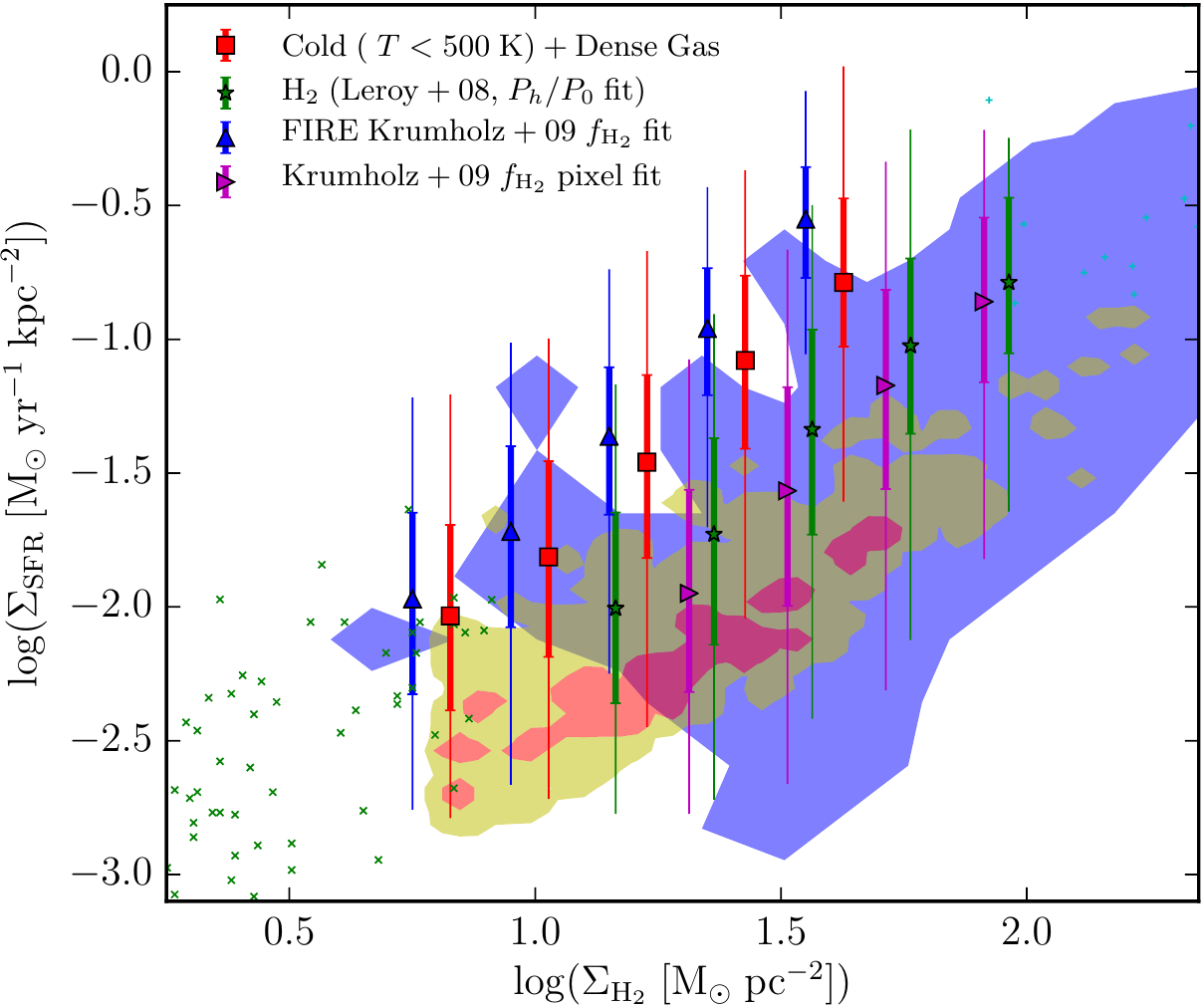}
   	\vspace{-0.25cm}
    	\caption{\rev{Comparison of three of proxies for the molecular gas surface density in the KS relation at 1~kpc$^2$ from a subset of the galaxy simulations presented in this work from $z \approx 0.2 - 0$.  Points, error bars, and shaded regions (molecular KS observations) are in the style of Fig.~\ref{tracers}.  The Cold \& Dense ($<300$ K and $>10$ cm$^{-3}$) tracer is calculated on a per-particle basis, whereas the H$_2$ masses predicted by an empirical fit from \citet{Leroy2008} and the molecular fraction fits of \citet{Krumholz2009b} are produced from kpc-averaged quantities of the mapped pixels, and applied directly to the gas particles themselves as is done in calculating the SFRs in FIRE.
%	The $f_{\rm H_2}$ values predicted by the fits from \citet{Leroy2008} and \citet{Krumholz2011} converge quickly to $\approx 1$ above 10 M$_\odot$~pc$^{-2}$, whereas the Cold \& Dense tracer 
%	suggesting that the FIRE simulations are producing correct SFRs for the pressure and large-scale properties of the ISM.  However, the $\sim 0.5$ dex discrepancy in gas surface density between those two empirical fits and the particle-scale Cold \& Dense gas tracer suggest that not enough gas is able to remain in or reach the highest resolvable densities in the simulations.  Other proxies for the molecular gas masses, including a warmer temperature cut (3000 K) or applying the \citet{Krumholz2011} fit directly to the particles, yield molecular fractions between the extremal cases presented here- implying a $\sim 0.5$~dex uncertainty in our estimates of molecular gas masses.
	The relations obtained when the \citet{Leroy2008} and \citet{Krumholz2009b} fitting functions are applied to predict the molecular gas surface density at the pixel scale are more consistent with observations than when the Cold \& Dense tracer is used, or when the \citet{Krumholz2009b} fit is applied at the particle scale, because at fixed SFR surface density, these tracers yield molecular gas surface densities $\sim 0.5$ dex lower than those obtained using the two aforementioned fitting functions. This result suggests that the FIRE simulations are producing `correct' SFRs given the large-scale properties of the ISM (mid-plane pressure and dust opacity), but insufficient gas is able to remain in or reach the highest resolvable densities in the simulations. Other proxies for the molecular gas mass, including a warmer temperature cut (3000 K), yield molecular fractions between the extremal cases presented here. These results suggest a $\sim 0.5$~dex uncertainty in our estimates of molecular gas surface density, with the Cold \& Dense tracer systematically biased low.
	}}
	\label{fig:CD_KG_compare}
\end{figure}
%%%% FIGURE B2: C&D to Neutral.
\begin{figure}
	\centering
	\includegraphics[width=0.47\textwidth]{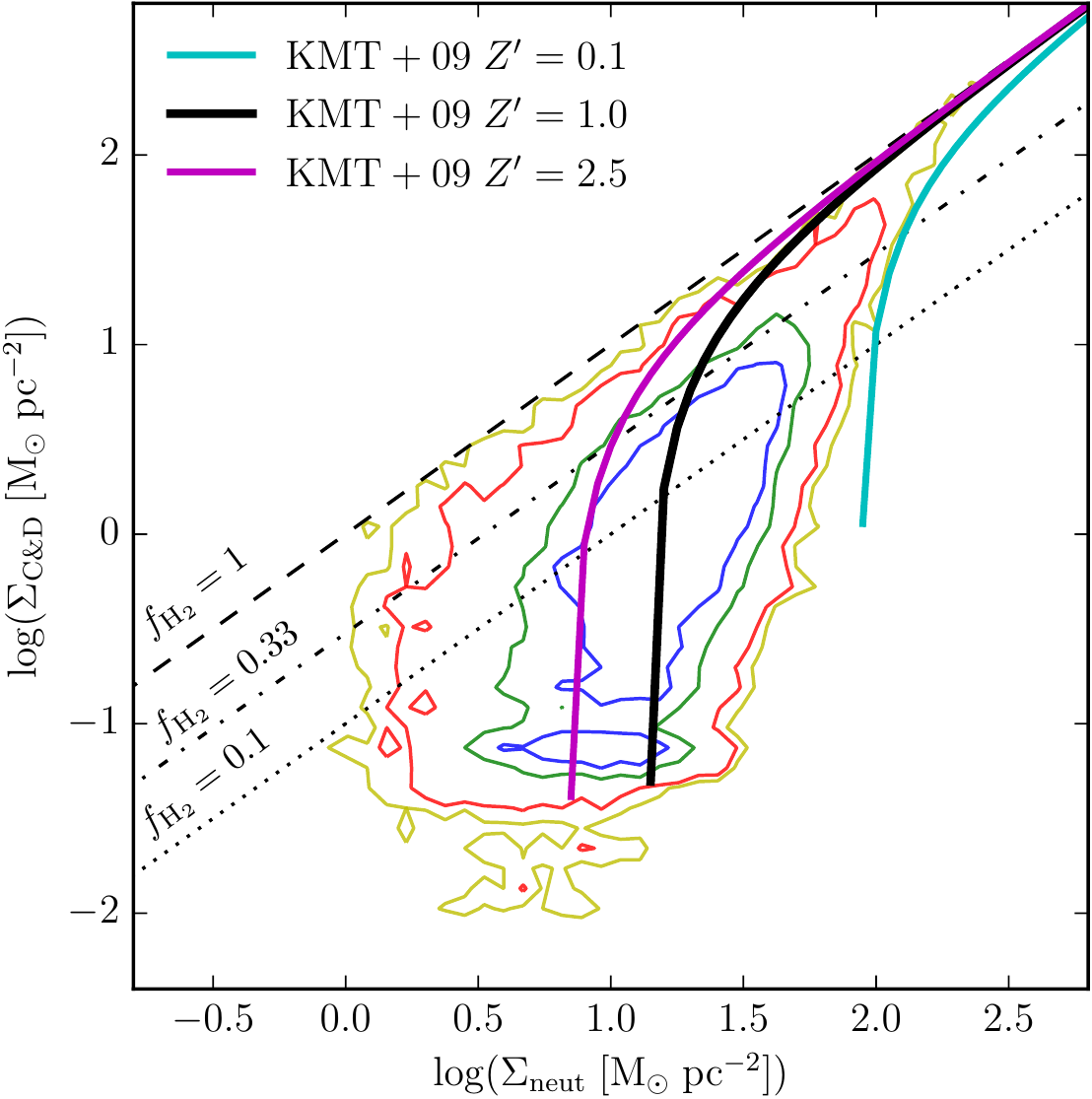}
   	\vspace{-0.25cm}
    	\caption{{Distribution of $\Sigma_{\rm C\& D}$, the Cold \& Dense gas surface density, versus $\Sigma_{\rm neut}$, the neutral gas surface density, for gas with metallicities $-0.1 < \log{Z/Z_\odot} < 0.1$ ($Z\approx Z_\odot$) in the FIRE simulations, with pixel sizes of 1~kpc.  Colored (yellow, red, green, blue) contours indicate (95, 90, 70, 50)th-percentile-inclusion contours of the data.  Black (dashed, dash-dotted, dotted) lines represents $f_{\rm H_2} = (1, 0.33, 0.1)$.  Colored lines (cyan, black, magenta) represent $f_{\rm H_2}$ molecular fraction fits for various metallicities ($Z/Z_\odot = 0.1,1.0,2.5$) from \citet{Krumholz2009b}. The core of the ``molecular fraction'' (as represented by $\Sigma_{\rm C\&D}/\Sigma_{\rm neut}$) has a steeply rising slope between $0.5 <\log{\Sigma_{\rm neut}} < 1.0$. However, the Cold \& Dense fraction does not converge to unity as quickly as the fits from \citet{Krumholz2009b} at solar metallicities and has a tail of high fractions to lower gas surface densities, below their metallicity-dependent thresholds.
    }\label{fig:molefrac}}
\end{figure}

% CCH continue here

\rev{As compared to the neutral gas (atomic \emph{and} molecular) mass, predicting the molecular mass in the FIRE simulations alone is difficult, as cooling and self-shielding are calculated using approximate look-up tables and are not done fully self-consistently with radiative transfer at the particle scale \citep[see][for details of the numerical implementation of cooling and shielding effects in FIRE-1]{Hopkins2014}.  For reasons presented in Appendix~\ref{sec:appendix:baryonic.physics}, getting the cooling and shielding even grossly incorrect in the coolest, densest gas generally has no \emph{dynamical} effect on the simulations but does greatly affect the high-density tail of the gas volumetric-density distribution.  Similarly, the gas density threshold for star formation and the instantaneous local star formation efficiency assumed affect the time that gas spends at the highest resolvable densities (see Figure~\ref{densCDFs}).  What truly constitutes molecular gas in the simulations, or what would be observed as such, requires careful forward modeling of the chemical abundances and molecular line emission.  That, coupled with large uncertainties in the gas phase structure at low temperatures and its dependence on the aforementioned grid/particle-scale star formation and cooling prescriptions leaves us with crude, though physically motivated, proxies for molecular gas masses in the simulations.  In this work, we have used a conservative estimator for molecular gas, our Cold \& Dense (T $< 300$ K {\em and} $n_{\rm H} > 10$ cm$^{-3}$) gas tracer.  We compare this explicitly with two other empirical estimators for molecular gas at the kpc-scale, and one other local (e.g., few pc) estimate, in Figure~\ref{fig:CD_KG_compare}.  There, we compare an estimator from \citet{Leroy2008}, for which the molecular fraction is taken to be directly proportional to the mid-plane pressure of the ISM, which is calculated using both the gas and stellar surface densities and dynamical times \citep[see][from which they adapt their empirical estimator]{Blitz2006}, and fits for the molecular fraction from \citet{Krumholz2009b} relating to local dust opacity applied both at the pixel (kpc) and at the particle (pc) scale.}

\rev{The two empirical estimators lie roughly 0.5~dex above the Cold \& Dense gas tracer at all gas surface densities and are in better concordance with observations, uncertainties in them notwithstanding (see Appendix~\ref{sec:appendix:xco}).  Due to the steepness of the \citet{Krumholz2009b} fitting function at the atomic-to-molecular transition, $\sim 10$ M$_\odot$ pc$^{-2}$, very few kpc-scale pixels contribute to the data shown (because many kpc-scale pixels have $\log\Sigma_{ \rm H_2} \ll -1$), indicating the necessity of assuming clumping factors when applying these fits on scales larger than GMCs themselves in low-gas surface density environments (e.g. disk outskirts).  This $\sim 0.5$~dex discrepancy indicates three things: (1) the FIRE simulations do appear to produce correct SFRs for the large-scale \emph{pressure} of the ISM, (2) the SFRs are in concordance with those expected given the large-scale optical depths of the ISM, and (3) the FIRE-1 simulations appear to either produce insufficient high-density gas or consume high-density gas more quickly than expected.  Points (1) and (2) lend credence to trusting the large-scale structure and dynamics of the ISM and the FIRE simulations; however, point (3) indicates that we have not yet converged on producing a realistic phase structure of the ISM \emph{at the highest densities near our resolution limits} \citep[noting that in FIRE-1 the gas density threshold for star formation is $\sim 50$ cm$^{-2}$, quite low compared to the densities of PDRs and the critical density of $^{12}$CO][]{Hollenbach1999}.  We compared several other estimators for the molecular fraction, including a less-stringent temperature cut ($T < 3000$ K) and a stellar surface density fit ($\Sigma_{\rm H_2} / \Sigma_{\rm HI}\propto \Sigma_\star$) also explored in \citet{Leroy2008}, but omit them for clarity as they all lay between the extremal values of the Cold \& Dense tracer at the low end and the \citet{Leroy2008}/\citet{Blitz2006} empirical pressure and \citet{Krumholz2009b} opacity fits at the high end.  Reconstructing the $f_{\rm H_2}$ fraction of the particles themselves using the fits of \citet{Krumholz2011} used in GIZMO (also seen in the figure), however agrees more closely (underestimating only by $\sim 0.1$~dex) with the Cold \& Dense gas tracer.  The fact that the same estimator applied at the pc- and kpc-scales can produce results with $0.5-0.7$~dex differences likely owes to point (3) and the difficulties in estimating the local column depths for shielding.  Throughout the main body of this paper, we use the Cold \& Dense tracer as a \emph{lower limit} on the molecular gas column and acknowledge a $\sim 0.5$~dex uncertainty in our dense gas tracer, dependent on our choice of proxy in order to most fairly show the range of tension between our results and observations given that choice.  In Figures~\ref{tracers}, \ref{tracers_eff}, \& \ref{res-dep}, we use arrows to indicate how shifting the molecular gas surface densities based on the Cold \& Dense tracer 0.5 dex higher would bring the simulations and observations in closer agreement.}

We explicitly compare the approximate molecular fraction versus neutral gas surface density relation obtained using the Cold \& Dense gas proxy (T $< 300$ K {\em and} $n_{\rm H} > 10$ cm$^{-3}$) to fits from \citet{Krumholz2009b}, with 1~kpc pixels.  A plot of $\Sigma_{\rm C\&D}$, the Cold \& Dense gas surface density, versus $\Sigma_{\rm neut}$, the total neutral gas (${\rm HI + H_2}$) surface density for gas with approximately solar metallicity ($Z_\odot \pm 0.1$~dex), is shown in Figure~\ref{fig:molefrac}.  Compared to the steep atomic-to-molecular transition thresholds found by \citet{Krumholz2009b}, the ratio $\Sigma_{\rm C\&D}/\Sigma_{\rm neut}$ converges much more slowly to unity (only near $\sim 100$ M$_\odot$ pc$^{-2}$).  The bulk of pixels ($\sim70$\%) lie below $f_{\rm H_2}=0.33$, thus indicating that the Cold \& Dense gas tracer is likely consistently underestimating the molecular fraction by $0.5-1$~dex for $\Sigma_{\rm gas} > 10$ M$_\odot$ pc$^{-2}$.  For gas surface densities between $1-10$ M$_\odot$ pc$^{-2}$, it is unclear how much of the high values for the molecular fractions at low gas surface densities is due to beam-filling (i.e. cloud-counting) effects, and it is unclear whether the Cold \& Dense gas tracer is over- or underestimating the molecular fractions there.

\section{Uncertainty in $X_{\rm CO}$ for Observed $\Sigma_{\rm H_2}$ and Tension with Simulations}
\label{sec:appendix:xco}
%%%%%%%%%%% FIGURE 2 - VARIOUS FORMULATIONS OF KS LAW PANEL
\begin{figure*}
	\centering
	\includegraphics[width=0.99\textwidth]{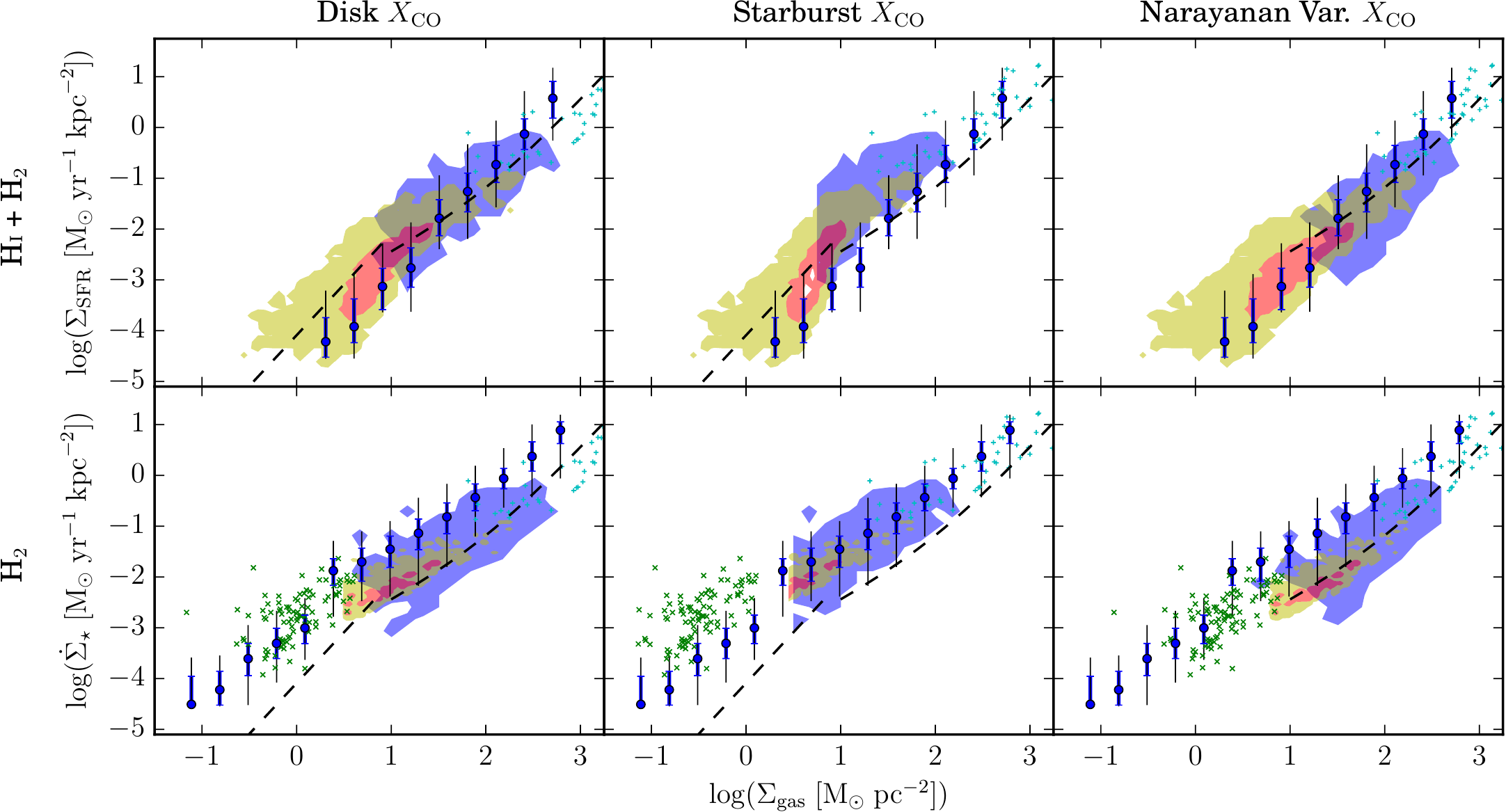}
	\caption{\rev{KS relation in the FIRE runs in 1 kpc$^2$ pixels, binned by $\Sigma_{\rm gas}$, for neutral and ``molecular" gas tracers (rows) and the 10~Myr-averaged star formation rate surface density, with three different $X_{\rm CO}$ conversion factors applied to observations (enumerated in Fig.~\ref{tracers}) for comparison (columns).  Atomic + molecular hydrogen is $\sim \Sigma_{\rm HI + H_2}$ (\emph{top row}), and Cold \& Dense gas includes particles with T $< 300$ K {\em and} $n_{\rm H} > 10$ cm$^{-3}$ ($\sim \Sigma_{\rm H_2}$, \emph{bottom row}).  All observations have been re-calibrated with either a standard ``star-forming disk" $X_{\rm CO} = 2 \times 10^{20}$~cm$^{-2}$/(K km s$^{-1}$) (\emph{left column}, as adopted by \citealt{Bigiel2008}), a ``starburst"  $X_{\rm CO} = X_{\rm CO, \, disk}/3.2$ (\emph{middle column}, adopted by \citealt{Genzel2010}), and a variable $X_{\rm CO}$ fit interpolating between the ``star-forming disk" and ``starburst" $X_{\rm CO}$ values (\emph{right column}, normalized by a factor of 2 higher than either of the other $X_{\rm CO}$'s) found by \citet{Narayanan2012}, as described in Section \ref{obsref}. The neutral gas observations have been decomposed into constituent $\Sigma_{\rm HI}$ and $\Sigma_{\rm H_2}$ columns, with the latter being corrected (see Section \ref{obsref}). Uncertainty in values of $X_{\rm CO}$, allowing for $\sim 0.5$~dex variations in observationally-inferred molecular gas masses can affect the \rev{(dis)agreement} between FIRE and observations, on the same order as variations in choices of molecular gas proxy in our mapping (see Fig.~\ref{fig:CD_KG_compare}). We adopt the \citet{Narayanan2012} variable $X_{\rm CO}$ interpolation function throughout the main text.}}
	\label{fig:varXco}
\end{figure*}

\rev{All of the observations to which we compare our results infer molecular hydrogen masses from CO emission using a single or bimodel CO-to-H$_2$ conversion factor, $X_{\rm CO}$, which is used to convert from CO linewidth ${\rm W}({\rm ^{12}C^{16}O} \, J = 1 \rightarrow 0)$ to H$_2$ column density ${\rm N(H_2)}$ using the following relation:
\be
{\rm N(H_2)} =X_{\rm CO} {\rm W}({\rm ^{12}C^{16}O} \, J = 1 \rightarrow 0) \, ,
\ee
The value of $X_{\rm CO}$ is on the order of $10^{20}$~cm$^{-2}$/(K km s$^{-1}$) \citep[see][for a review on the $X_{\rm CO}$ conversion factor]{Bolatto2013},
but there is tremendous disagreement about the exact value it takes and its dependences on surface density, metallicity, and other parameters.  As a result, we find it necessary to understand the extent to which the observational data can vary for differing, but reasonable, assumptions about $X_{\rm CO}$.  Figure~\ref{fig:varXco} shows how various choices for the value of $X_{\rm CO}$ affect the tension between observations and our ``standard" tracers of atomic + molecular ($\Sigma_{\rm HI + H_2}$) and Cold \& Dense ($\sim \Sigma_{\rm H_2}$) gas surface density.  We compare three conversion factors: (\emph{left column}) a ``star-forming disk" $X_{\rm CO} = 2 \times 10^{20}$ cm$^{-2}$/(K km s$^{-1}$), a value widely adopted for low-redshift observations of Milky-Way like galaxies \citep{Strong1996, Dame2001,Bigiel2008, Genzel2010, Shapiro2010, Wei2010, Tacconi2013, Amorin2016}; (\emph{middle column}) a ``starburst" $X_{\rm CO} = X_{\rm CO, \, disk}/3.2$ which is a factor of 3.2 smaller than the disk conversion factor, owing to the fact that at high gas surface densities in extreme star-forming systems the disk $X_{\rm CO}$ predicts gas masses in excess of observed dynamical masses, which is a known problem for ULIRG observations \citep{Solomon1997, Downes1998, Solomon2005, Bothwell2010}; and (\emph{right column}) a variable $X_{\rm CO}$ interpolation function based on \citet{Narayanan2012}.  We take the form of the \citet{Narayanan2012} interpolation function to be
\begin{equation} \label{eq:N12_xco}
X_{\rm CO} = min[4, 6.75 \times W_{\rm CO}^{-0.32}] \times 10^{20} \frac{\rm cm^{-2}}{\rm K \, km \, s^{-1}} \; ,
\end{equation}
which is identical to that presented in their work\footnote{It is noted that their normalization/maximum $X_{\rm CO}$ is twice that of the ``star-forming disk" $X_{\rm CO}$ factor.}, assuming a Solar gas metallicity \citep[see][for a comparable interpolation function]{Ostriker2011}.  We recalibrate all of the observations enumerated in Section \ref{obsref} for the KS relation using the \xco value predicted using Eq. \eqref{eq:N12_xco}.  To correct the $\Sigma_{\rm H{\scriptsize I} +H_2}$ measurements, we decomposed the total column into atomic and molecular components (the latter then being corrected in the manner of the $\Sigma_{\rm H_2}$ values) using data from the references themselves, where available, or assuming a molecular fraction fit from \citet{Leroy2008} when necessary.}

\rev{In Appendix~\ref{sec:appendix:molefrac}, we demonstrated that the ratio of Cold \& Dense tracer to the neutral hydrogen surface density slowly converges to one above $\sim 10$~M$_\odot$ pc$^{-2}$ and hovers $\sim 0.5$~dex below other empirical fits and the neutral gas surface density in the KS plane until $\gtrsim 100$~M$_\odot$ pc$^{-2}$.  Considering this, adopting the disk \xco leads to large disagreement at the highest gas surface densities, which we believe our molecular gas surface density proxy is nearly converged for \citep[and other studies of the FIRE simulations have shown that the centers of our Milky Way-mass galaxies are not outliers in terms of gas surface density or star formation rate][]{Hopkins2014,Hopkins2016, Torrey2016}, whereas a purely ``starburst" \xco seems to suggest our simulations are over-predicting neutral gas surface densities by $\sim 0.5$~dex everywhere but at the most extreme gas surface densities.  Given that there is little support for either of these values of \xco holding for all gas surface densities, it is reasonable to use an interpolation function, such as that of \citet{Narayanan2012}, for the range of observations.}
%Though the normalization of their interpolation function for \xco differs by a factor of two from the disk \xco usually adopted, reasonable agreement is seen both at the ``starburst" end of the gas surface density distribution and between the data and simulations in normal disk conditions.}
 
\rev{Otherwise, between the disk and starburst \xco factors a $\sim 0.5$~dex uncertainty exists, before even considering reasonable additional factor of a few differences in those values themselves \citep{Bolatto2013}.  This level of variation is on the order of the difference between extremal estimators of our molecular gas masses (see the difference between the Cold \& Dense tracer and the \citealt{Krumholz2009b} relation applied to individual pixels shown in Figure~\ref{fig:CD_KG_compare}). Although the Cold \& Dense gas tracer is clearly a conservative estimate of the molecular gas mass in the simulations and more careful forward-modeling of CO emission is clearly necessary (motivating a future work), the uncertainty in the observational value of \xco makes it difficult to determine the absolute level of (dis)agreement between observations and simulations (all simulations, not just the FIRE simulations) at the $\sim 0.5-1$~dex level.  As a result, in order to attempt to compare the results from FIRE on an appropriate footing with the observations throughout this paper, which both cover a parameter space of star formation rates and gas surface densities in both the ``star-forming disk" and ``starburst" regimes, we recalibrate the compiled observations with the \citet{Narayanan2012} \xco interpolation function.}
%%%%%%
\label{lastpage}
\end{document}